\documentclass[aps,prx,10pt,twocolumn,superscriptaddress,longbibliography,nofootinbib,floatfix]{revtex4-2}

\usepackage{amsmath,amssymb,mathtools,bm} 
\usepackage{amsthm}
\usepackage{graphicx}
\usepackage{booktabs}
\usepackage{microtype}
\usepackage{xcolor}
\usepackage{hyperref}
\hypersetup{colorlinks=true,allcolors=blue}
\graphicspath{{figures/}}

\newcommand{\FiniteHill}{0.633}
\newcommand{\FinitePair}{0.339}
\newcommand{\DistanceSmall}{0.109}
\newcommand{\DistanceLarge}{0.0100}
\newcommand{\ContourExponent}{1.15}
\newcommand{\ContourExponentSE}{0.02}
\newcommand{\SingleCHill}{0.04880}
\newcommand{\SingleCPair}{0.01957}
\newcommand{\SingleBHill}{0.00811}
\newcommand{\SingleCRatio}{2.49}
\newcommand{\IsolatedHill}{0.623}
\newcommand{\IsolatedPair}{0.324}
\newcommand{\IsolatedGapHill}{0.010}
\newcommand{\IsolatedGapPair}{0.015}
\newcommand{\TargetSemiC}{5.97}
\newcommand{\TargetSemiB}{11.94}
\newcommand{\OmegaEpsilon}{1.6\times10^{-4}}
\newcommand{\SourceConcentration}{0.00125}
\newcommand{\SourceFoldLow}{1.412963}
\newcommand{\SourceFoldHigh}{5.848452}
\newcommand{\SourceCLow}{0.001263}
\newcommand{\SourceBLow}{0.002526}
\newcommand{\SourceCSaddle}{0.132519}
\newcommand{\SourceBSaddle}{0.265037}
\newcommand{\SourceCHigh}{1.867469}
\newcommand{\SourceBHigh}{3.734937}
\newcommand{\PairDriftFold}{1.43614}
\newcommand{\PairDriftFoldShift}{0.02193}

\newcommand{\PairDriftB}{1.0625}

\newcommand{\LossCSymmetric}{0.2946}

\newcommand{\LossCQuarter}{0.3572}

\newcommand{\LossBQuarter}{0.8372}
\newcommand{\LossBFour}{0.3254}
\newcommand{\ControlHillReturnHeavy}{0.263}

\newcommand{\ControlHillFeedforwardHeavy}{0.921}
\newcommand{\ControlPairReturnHeavy}{0.066}

\newcommand{\ControlPairFeedforwardHeavy}{0.799}
\newcommand{\GammaFiftyHill}{1.900}
\newcommand{\GammaFiftyPair}{2.148}
\newcommand{\LawShiftEtaTwo}{0.248}
\newcommand{\HillEtaHalfShift}{0.215}
\newcommand{\HillEtaDoubleShift}{0.151}
\newcommand{\PairEtaHalfShift}{0.286}
\newcommand{\PairEtaDoubleShift}{0.205}
\newcommand{\HillOwnFoldExcess}{0.48568}
\newcommand{\PairOwnFoldExcess}{0.71200}
\newcommand{\OwnFoldResidualGap}{0.22632}
\newcommand{\ControlSourceCriticalR}{0.06353}
\newcommand{\ControlSourceCriticalGammaB}{50.82}

\newtheorem{proposition}{Proposition}
\newtheorem{theorem}{Theorem}
\newcommand{\dd}{\mathrm{d}}
\newcommand{\Ree}{\operatorname{Re}}
\newcommand{\spec}{\operatorname{spec}}
\newcommand{\al}{\alpha}
\newcommand{\E}{\mathbb{E}}
\newcommand{\Prob}{\mathbb{P}}
\newcommand{\Rplus}{\mathbb{R}_{\geq 0}}
\newcommand{\Borg}{\mathcal{B}_{+}}
\newcommand{\Aorg}{\mathcal{A}_{+}}

\begin{document}

\title{Finite-amplitude establishment beyond local invasion in cooperative chemical networks}

\author{Ferran Mart\'inez-Reyes}
\affiliation{Independent Researcher, Bigues i Riells del Fai, Catalonia, Spain}

\author{Slava G. Turyshev}
\affiliation{Jet Propulsion Laboratory, California Institute of Technology, 4800 Oak Grove Drive, Pasadena, California 91109, USA}

\date{\today}

\begin{abstract}
Local kinetic exponents and next-generation radii diagnose infinitesimal invasion, but they need not determine whether a nonlinear organized state is accessible. Motivated by the kinetic component of Gate~7 in a twelve-gate nonequilibrium framework, we study cooperative networks whose productive linearization contains no closed cycle. We prove that an acyclic productive Jacobian has $R=0$ and $\lambda_{\rm kin}=-\min_i d_i<0$ for every drive, while a nonlinear return path can create a stable finite-amplitude branch through a saddle-node. In a canonical catalyst--assembly model, the saddle's stable manifold---not a threshold in either concentration---is the deterministic establishment boundary, so equal-abundance seeds of different composition can have opposite outcomes. At finite copy number, establishment becomes a committor and first-passage problem. For a benchmark protocol with $\Gamma=2$, $\Omega=16$, $\eta=0.02$, and horizon $T=1000$, density-Hill and pair-count generators give first-entrance probabilities $0.633$ and $0.339$; their interpolated half-success drives are $1.900$ and $2.148$, an approximately $13.1\%$ shift. The associated constant-source deterministic system retains a stable low state, so these transitions are fluctuation- and discrete-arrival-enabled. The finite-amplitude regime is not confined to exact acyclicity: a weak linear return leaves a finite interval with $0<R<1$, $\lambda_{\rm kin}<0$, and bistability. A general cooperative-return criterion separates the superlinear local condition from the global return-profile structure required for a fold. The resulting hierarchy of branch existence, basin accessibility, and finite-time stochastic establishment replaces a single onset threshold by a protocol-dependent kinetic criterion.
\end{abstract}

\maketitle

\section{Introduction}

An autocatalytic chemical system may maintain production after inoculation yet fail to start under the same feed and loss conditions. The distinction is relevant to self-starting reactors, catalytic organization, protocell models, and prebiotic chemistry: existence of a maintained state does not determine whether the material initially available can reach it.

This question arose from Gate~7 of the twelve-gate nonequilibrium framework introduced by one of us (SGT) \cite{Turyshev2026}. Gate~7 combines three logically distinct requirements: structural and stoichiometric closure, including a sufficiently connected catalysis relation; dissipative thermodynamic feasibility; and kinetic accessibility against loss. The present paper addresses only the third. The local convention is unchanged from Ref.~\cite{Turyshev2026}: the productive Jacobian is separated from the loss operator, the extinction-state exponent is the spectral abscissa of their difference, and the next-generation quantity is a dimensionless spectral radius. Thus the phenomenon below is not a definitional mismatch. It shows instead that a positive local exponent is sufficient for infinitesimal invasion but is not a mechanism-independent necessary condition for finite-amplitude establishment. The dilute-state linearization is therefore a scope assumption defining the continuous-invasion branch, not a complete characterization of kinetic accessibility. The structural and thermodynamic legs of Gate~7 remain separate tests. At finite abundance the reduced dependency graph studied below closes as $C\to B\to C$, but that dynamical feedback statement is not by itself a proof of RAF membership or stoichiometric autocatalysis for a specified elementary chemical reaction system.

Initiation and maintenance are already distinguished in chemical bistability, Allee effects, and autocatalytic-network models \cite{Schlogl1972,Dennis1989,OhtsukiNowak2009,GiriJain2012}. Structural descriptions identify autocatalytic sets and cores \cite{HordijkSteel2004,Blokhuis2020}, including networks activated by introduced species and subsequently maintained by external food \cite{Peng2022}. Experiments on autocatalytic bistability, cooperative replicators, and model protocells further demonstrate that preparation and composition can matter \cite{Semenov2016,LincolnJoyce2009,Vaidya2012,AdamalaSzostak2013}.

Two descriptions are then required beyond local stability. Deterministically, the successful preparations form a basin whose boundary is generally a codimension-one invariant set, not a threshold in a single concentration \cite{Menck2013}. At molecular copy number, success is a first-passage event for a specified jump process \cite{EVandenEijnden2006,Metzner2009}; low-immigration colonization under an Allee effect illustrates the role of rare arrivals \cite{Beer2015}, while stochastic reductions show why a deterministic Hill law alone does not determine few-molecule fluctuations \cite{KimJosicBennett2014}.

The remainder of this paper is organized as follows. Section~\ref{sec:local_theory} develops the local-invasion diagnostics and cooperative-return criterion. Section~\ref{sec:canonical} analyzes the canonical catalyst--assembly model, including its saddle-node and stable-manifold basin boundary. Section~\ref{sec:stochastic} formulates finite-copy establishment through microscopic reaction laws, committors, sources, and first-passage times. Section~\ref{sec:general_theory} tests robustness to controls, losses, cooperative order, network depth, and weak linear return. Section~\ref{sec:discussion} synthesizes branch existence, basin accessibility, and stochastic establishment and applies this distinction to Gate~7. Appendices~\ref{app:proofs}--\ref{app:numerics} contain the proofs, mass-action realization, and numerical protocols.

\section{Local invasion and nonlinear closure}
\label{sec:local_theory}

Consider an open reaction system
\begin{equation}
 \dot{\bm{x}}=\bm{s}+\bm{F}(\bm{x};\Gamma)-\Lambda\bm{x},
 \qquad \bm{x}\in\Rplus^m,
 \label{eq:general}
\end{equation}
where $\bm{s}$ is independent of the internal species, $\bm{F}$ contains productive reactions, and $\Lambda=\operatorname{diag}(d_1,\ldots,d_m)$ with $d_i>0$ collects first-order losses. If $\bm{s}=0$ and $\bm{F}(\bm{0};\Gamma)=0$, extinction is an equilibrium. Define
\begin{align}
 J_{\rm prod}(\Gamma)&=D_{\bm{x}}\bm{F}(\bm{0};\Gamma),\label{eq:Jprod}\\
 A_0(\Gamma)&=J_{\rm prod}(\Gamma)-\Lambda,\label{eq:A0}\\
 \lambda_{\rm kin}(\Gamma)&=\al(A_0)
 =\max_{\lambda\in\spec(A_0)}\Ree\lambda,\label{eq:lkin}\\
 K_0(\Gamma)&=J_{\rm prod}(\Gamma)\Lambda^{-1},\qquad
 R(\Gamma)=\rho(K_0).\label{eq:K0}
\end{align}
Here $J_{\rm prod}$ excludes the explicitly separated losses, $A_0$ is the full extinction-state Jacobian, $\lambda_{\rm kin}$ has units of inverse time, and $R$ is dimensionless. Because $\Lambda$ is invertible, $J_{\rm prod}\Lambda^{-1}$ and $\Lambda^{-1}J_{\rm prod}$ are similar and have the same spectral radius. For $J_{\rm prod}\geq0$, positive-system theory gives
\begin{equation}
 \lambda_{\rm kin}>0 \quad\Longleftrightarrow\quad R>1,
 \label{eq:threshold_equiv}
\end{equation}
without implying numerical equality \cite{Thieme2009,FarinaRinaldi2000}. If $\bm{s}\neq0$, exact zero is not an equilibrium and these quantities must instead be evaluated about the source-maintained background. These definitions separate source-driven onset, local invasion of an unstable extinction state, and finite-amplitude establishment in which extinction remains stable but a distinct organized attractor is reachable from a nonzero preparation.

\begin{proposition}[Acyclic productive linearization]
\label{prop:acyclic}
If $J_{\rm prod}\geq0$ and its directed graph is acyclic, then
\begin{equation}
 R=0,\qquad \lambda_{\rm kin}=-\min_i d_i<0,
 \label{eq:acyclic_result}
\end{equation}
independently of the magnitudes of the nonzero feed-forward entries.
\end{proposition}
A proof is given in Appendix~\ref{app:proofs}. The proposition states that arbitrarily strong feed-forward production can leave every local invasion diagnostic subcritical when no productive cycle closes at linear order. It does not by itself imply a finite-amplitude state; that requires global nonlinear feedback.

A minimal class that supplies such feedback is the $n$-stage cooperative-return chain
\begin{subequations}
\label{eq:nchain}
\begin{align}
 \dot{x}_1 &= a_1\Gamma\,\phi(x_n)-d_1x_1,\label{eq:nchain1}\\
 \dot{x}_i &= a_i\Gamma x_{i-1}-d_ix_i,
 \qquad i=2,\ldots,n.\label{eq:nchaini}
\end{align}
\end{subequations}
Assume $n\geq2$, $\phi(0)=\phi'(0)=0$, and $\phi(x)>0$ for $x>0$. With all $x_i$ expressed as concentrations and $\phi$ dimensionless, $a_1$ has units of concentration per time and $a_i$ for $i\geq2$ have units of inverse time. This scalar return-profile formulation is a specialized explicit instance of established monotone positive-feedback multistability criteria \cite{AngeliSontag2004,AngeliFerrell2004}. Let
\begin{equation}
 \chi(x)=\frac{\phi(x)}{x},\qquad
 A=\prod_{i=1}^{n}a_i,
 \qquad D=\prod_{i=1}^{n}d_i.
 \label{eq:qdef}
\end{equation}

\begin{theorem}[Cooperative-return fold]
\label{thm:fold}
Suppose $\phi\in C^1([0,\infty))\cap C^2((0,\infty))$ is nondecreasing, $\chi(x)\to0$ as $x\to0^+$ and $x\to\infty$, and $\chi$ has a unique critical point $x_{\rm sn}>0$, with $\chi' >0$ on $(0,x_{\rm sn})$, $\chi'<0$ on $(x_{\rm sn},\infty)$, and $\chi''(x_{\rm sn})<0$. Write $\chi_{\max}=\chi(x_{\rm sn})$. Then extinction is exponentially stable for every $\Gamma$, with $R=0$ and $\lambda_{\rm kin}=-\min_i d_i$. Positive equilibria satisfy
\begin{equation}
 \chi(x_n^*)=\frac{D}{A\Gamma^n},
 \label{eq:qintersection}
\end{equation}
and are created in a saddle-node at
\begin{equation}
 \Gamma_{\rm sn}=\left(\frac{D}{A\chi_{\max}}\right)^{1/n}.
 \label{eq:Gsn_general}
\end{equation}
At a positive equilibrium, define the logarithmic return slope
\begin{equation}
 \nu_\phi(x_n^*)=\frac{x_n^*\phi'(x_n^*)}{\phi(x_n^*)}.
 \label{eq:elasticity}
\end{equation}
The equilibrium is linearly stable for $\nu_\phi<1$ and unstable for $\nu_\phi>1$; equivalently, the branch with $\chi'<0$ is stable and that with $\chi'>0$ is unstable.
\end{theorem}
Appendix~\ref{app:chain} gives the proof and consequences. The local condition $\phi(x)\sim cx^p$ with $p>1$ ensures $\phi'(0)=0$ and hence the absence of a linear return edge. It does not ensure a fold: finite-amplitude establishment additionally requires the global return profile $\chi=\phi/x$ to rise and then fall, as occurs for saturating cooperative responses.

\section{Canonical model and deterministic accessibility}
\label{sec:canonical}

\subsection{Cooperative maintenance}

The minimal two-stage, Hill-2 instance of Theorem~\ref{thm:fold} is an open, well-mixed system with buffered precursors,
\begin{subequations}\label{eq:canonical}
\begin{align}
 \dot C&=a\Gamma\frac{B^2}{K^2+B^2}-d_CC,\label{eq:Cdot}\\
 \dot B&=b\Gamma C-d_BB.\label{eq:Bdot}
\end{align}
\end{subequations}
The catalyst $C$ produces assembly-forming component $B$, which cooperatively enables catalyst production. Both variables are concentrations. The parameters $a$, $b$, and $d_C,d_B$ have units of concentration/time, inverse time, and inverse time, respectively; $K$ is a concentration. The dimensionless control $\Gamma$ multiplies both productive legs along a convenient diagonal path. It is not a chemical affinity or input power; independent controls are treated in Sec.~\ref{sec:control_path}. Assembly geometry and compartment growth are outside the state description. Appendix~\ref{app:massaction} gives a dimerization--binding realization of the deterministic return and states the reduction conditions.

At extinction, the productive and full Jacobians are
\begin{equation}
 J_{\rm prod}=\begin{pmatrix}0&0\\b\Gamma&0\end{pmatrix},\qquad
 A_0=\begin{pmatrix}-d_C&0\\b\Gamma&-d_B\end{pmatrix}.
 \label{eq:canonical_matrices}
\end{equation}
Thus the local growth exponent and next-generation radius are
\begin{equation}
 \lambda_{\rm kin}=\max\Ree\spec(A_0)=-\min(d_C,d_B),\qquad R=0.
 \label{eq:canonical_local}
\end{equation}
The productive Jacobian is nonzero but nilpotent: the linear edge $C\to B$ remains, while the cooperative return $B\to C$ has zero derivative at extinction. At finite abundance the two dependencies close a dynamical feedback loop. Equation~\eqref{eq:canonical_local} therefore agrees with both local conventions in Sec.~\ref{sec:local_theory}; strict local decay does not exclude a separate finite-amplitude attractor.

Writing $\mu=ab\Gamma^2/(Kd_Cd_B)$, the positive equilibria are
\begin{equation}
 B_\pm=\frac K2\left(\mu\pm\sqrt{\mu^2-4}\right),\qquad
 C_\pm=\frac{d_B}{b\Gamma}B_\pm,
 \label{eq:eqbranches}
\end{equation}
with fold
\begin{equation}
 \Gamma_{\rm sn}=\sqrt{\frac{2Kd_Cd_B}{ab}},\qquad B_{\rm sn}=K.
 \label{eq:Gsn_canonical}
\end{equation}
The positive-state Jacobian has trace $-(d_C+d_B)$ and determinant
\begin{equation}
 \det A_*=d_Cd_B\frac{B^2-K^2}{B^2+K^2}.
 \label{eq:detcanonical}
\end{equation}
The lower branch is a saddle and the upper branch is stable. Extinction remains stable, so $\Gamma>\Gamma_{\rm sn}$ gives bistability between extinction and the organized state. A Lyapunov argument proves boundedness and convergence to equilibria in the physical quadrant (Appendix~\ref{app:proofs}).

Unless stated otherwise, figures use reference concentration and loss-time units with $a=b=d_C=d_B=K=1$, so $\Gamma_{\rm sn}=\sqrt2$. Seeds are specified in these fixed $(C,B)$ coordinates as $\Gamma$ varies. The molecular scale is $\Omega=VN_Ac_{\rm ref}$ for volume $V$ and reference molar concentration $c_{\rm ref}$; $N_C/\Omega$ and $N_B/\Omega$ are normalized concentrations. Appendix~\ref{app:proofs} gives the full nondimensionalization.

\begin{figure*}[t]
\includegraphics{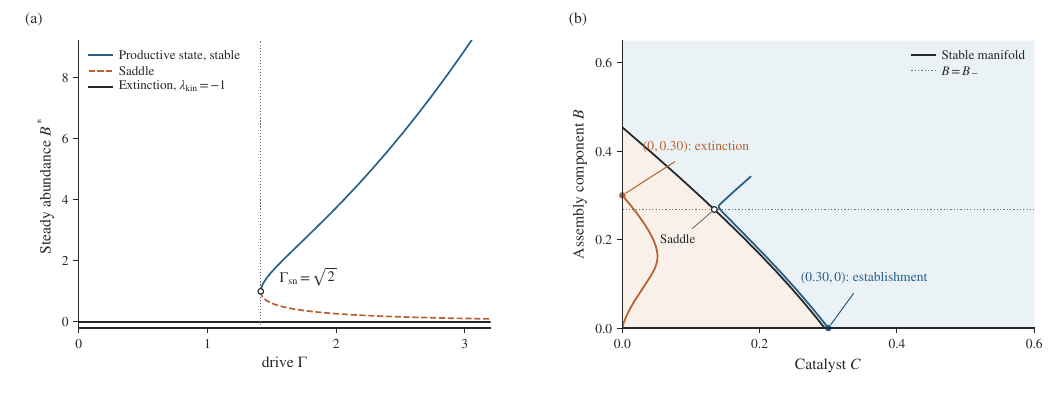}
\caption{Canonical finite-amplitude onset for $a=b=d_C=d_B=K=1$. (a) Extinction is stable for all $\Gamma$. A saddle and a stable organized branch are created at $\Gamma_{\rm sn}=\sqrt{2}$. (b) At $\Gamma=2$, the saddle's stable manifold separates extinction (orange) from establishment (blue). The horizontal level $B=B_-$ is not the boundary: $(C,B)=(0,0.30)$ returns to extinction although $B>B_-$, whereas $(0.30,0)$ establishes with no initial $B$. Numerical protocols are in Appendix~\ref{app:numerics}.}
\label{fig:canonical}
\end{figure*}

\subsection{Why seed composition matters}
\label{sec:basin}

The saddle stable manifold $W^s(C_-,B_-)$ separates the two basins. At $\Gamma=2$, Fig.~\ref{fig:canonical}(b) shows equal-total-abundance seeds with opposite outcomes: $(C,B)=(0,0.30)$ decays although $B>B_-$, whereas $(0.30,0)$ establishes. The saddle concentration $B_-$ is therefore not a one-coordinate seeding threshold. Assembly-rich material can disappear before generating sufficient catalyst; a catalyst-rich seed can produce the material needed to activate the cooperative return.

The local feed-forward response helps explain this asymmetry. For equal losses $d_C=d_B=d$, a pure catalyst seed has linearized response
\begin{equation}
 B_{\rm lin}(t)=b\Gamma C_0te^{-dt},\qquad
 \max_t B_{\rm lin}(t)=\frac{b\Gamma C_0}{ed}.
 \label{eq:transient_gain}
\end{equation}
Increasing $\Gamma$ amplifies the transient response while leaving the extinction eigenvalues fixed, an instance of nonnormal growth \cite{Trefethen1993}. This explains how a seed already in the productive basin can activate feedback; autonomous deterministic trajectories remain in their initial basin.

\begin{figure*}[t]
\includegraphics{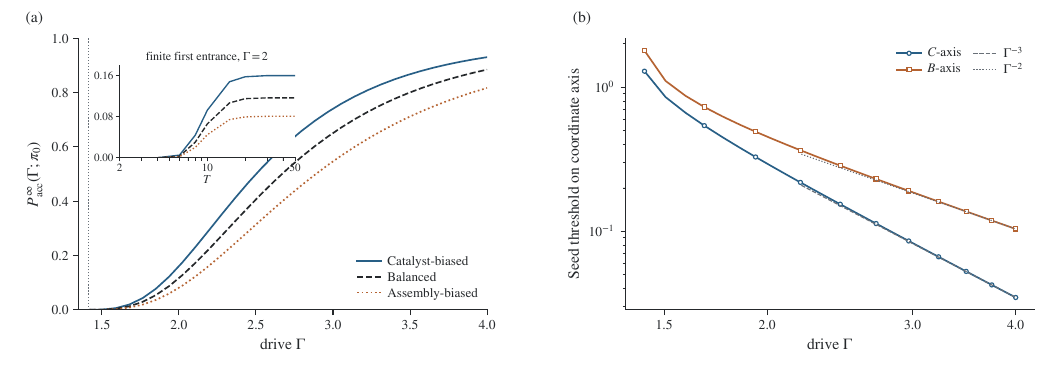}
\caption{Seed-dependent deterministic accessibility in the normalized canonical model. (a) Basin-weighted asymptotic accessibility for three composition distributions with the same total-seed distribution. Curves average eight independent scrambled Sobol replicates of $2^{16}$ points each; bands quantify numerical quadrature uncertainty. The inset shows finite first-entrance probabilities at $\Gamma=2$ for the same three preparations and the organized target used in the stochastic analysis. (b) Stable-manifold intercepts on the coordinate axes. Gray guides show the high-drive scalings obtained from Eq.~\eqref{eq:rescaled_highG}.}
\label{fig:detprob}
\end{figure*}

For a normalized seed distribution $\pi_0$, asymptotic deterministic accessibility is its organized-basin weight,
\begin{equation}
 P_{\rm acc}^{\infty}(\Gamma;\pi_0)
 =\int_{\Borg(\Gamma)}\pi_0(C,B)\,\dd C\,\dd B.
 \label{eq:Pdet}
\end{equation}
For a finite environmental window, let $\tau_+(\bm x_0)$ be the first entrance time into an organized target $\Aorg$ and define
\begin{equation}
 P_{\rm est}^{\rm det}(T;\Gamma,\pi_0)
 =\int \mathbf 1\!\left\{\tau_+(\bm x_0)\leq T\right\}\pi_0(\dd\bm x_0).
 \label{eq:Pdet_finite}
\end{equation}
Under the global convergence result in Appendix~\ref{app:proofs}, $\lim_{T\to\infty}P_{\rm est}^{\rm det}(T;\Gamma,\pi_0)$ equals Eq.~\eqref{eq:Pdet} provided that $\Aorg\subset\Borg(\Gamma)$ and $\Aorg$ contains a neighborhood of the stable organized equilibrium. A relative-radius target can cross the separatrix sufficiently close to the fold; in that case Eq.~\eqref{eq:Pdet_finite} remains a first-entrance measure but its infinite-time limit need not equal the basin weight. The target used for the $\Gamma=2$ inset is numerically contained in the organized basin. Figure~\ref{fig:detprob}(a) uses Eq.~\eqref{eq:Pdet} for asymptotic accessibility and holds the total seed $S$ fixed in distribution while varying its composition: $S\sim\mathrm{Gamma}(2,0.10)$, $f\sim\mathrm{Beta}(\alpha,\beta)$, and $(C_0,B_0)=(fS,(1-f)S)$. The Gamma distribution uses shape $2$ and scale $0.10$, giving $\E S=0.20$. The catalyst-biased, balanced, and assembly-biased protocols use $(\alpha,\beta)=(5,1.5),(2,2),(1.5,5)$. These illustrative preparations have identical total-seed statistics but different asymptotic accessibility probabilities. Their uncertainty bands measure numerical quadrature error; the assumed preparation distributions are not measured prebiotic statistics, and $S$ is an abundance budget rather than an energetic cost.

The seed boundary also moves anisotropically with drive. Near extinction, set $u=\Gamma^3C$ and $v=\Gamma^2B$ in the normalized model:
\begin{equation}
 \dot u=\frac{v^2}{1+v^2/\Gamma^4}-u,\qquad \dot v=u-v.
 \label{eq:rescaled_highG}
\end{equation}
The limiting phase portrait is independent of $\Gamma$, giving $C$- and $B$-axis intercepts proportional to $\Gamma^{-3}$ and $\Gamma^{-2}$ [Fig.~\ref{fig:detprob}(b)]. Composition therefore remains a relevant control even when increasing drive makes the required seed small.

\section{Molecular events and finite-time establishment}
\label{sec:stochastic}

\subsection{Reaction eligibility at low copy number}

For counts $(N_C,N_B)$, consider the normalized density-Hill jump process
\begin{align}
 w_{C,H}^+&=\Omega \Gamma\frac{N_B^2}{\Omega^2+N_B^2},& w_C^-&=N_C,\label{eq:ratesC}\\
 w_B^+&=\Gamma N_C,&w_B^-&=N_B.\label{eq:ratesB}
\end{align}
It has Eq.~\eqref{eq:canonical} as its deterministic concentration limit \cite{Kurtz1972,Gillespie1977,Gillespie2000}. To isolate the role of molecular eligibility, change only the return to
\begin{equation}
 w_{C,P}^+=\Omega \Gamma\frac{N_B(N_B-1)}{\Omega^2+N_B(N_B-1)}.
 \label{eq:pair_rate}
\end{equation}
The pair-count law requires two $B$ molecules; the ordered-pair convention absorbs the usual factor of two into the coefficient. Both laws are effective stochastic models. The pair-count condition is compatible with a dimer requirement, but neither law is derived here by eliminating the binding intermediates of Appendix~\ref{app:massaction}. For the deterministic reduction parameters used there, $\Omega\varepsilon=\OmegaEpsilon$ at $\Omega=16$, far below one site per compartment; the present jump laws must therefore be interpreted as alternative low-copy coarse grainings, not as integer-site realizations of that mechanism.

At fixed positive concentration $B=N_B/\Omega$, the laws have the same macroscopic limit. At low count they differ appreciably: for $n=N_B\geq1$,
\begin{equation}
 \frac{w_{C,H}^+-w_{C,P}^+}{w_{C,H}^+}
 =\frac{\Omega^2}{n[\Omega^2+n(n-1)]}.
 \label{eq:relative_eligibility}
\end{equation}
This relative discrepancy tends to zero at fixed $B>0$, but to $1/n$ at fixed $n$; it is exactly one for a singleton. The absolute rate difference is
\begin{equation}
 w_{C,H}^+-w_{C,P}^+
 =\frac{\Omega^3\Gamma n}
 {\left(\Omega^2+n^2\right)\left[\Omega^2+n(n-1)\right]}.
 \label{eq:absolute_eligibility}
\end{equation}
Writing $u=\Omega^{-1}$ and using the algebraic family $\phi_\ell(B)=(\ell B+B^2)/(1+\ell B+B^2)$ introduced in Eq.~\eqref{eq:leaky_return}, the two concentration drifts are exactly
\begin{equation}
 \frac{w_{C,H}^+}{\Omega}=\Gamma\phi_0(B),\qquad
 \frac{w_{C,P}^+}{\Omega}=\Gamma\phi_{-u}(B).
 \label{eq:exact_drift_family}
\end{equation}
Consequently, the drift difference satisfies the exact identity
\begin{equation}
\begin{aligned}
 \Delta f_\Omega(B)
 &=\Gamma\bigl[\phi_0(B)-\phi_{-u}(B)\bigr]\\
 &=\frac{\Gamma uB}{(1+B^2)(1+B^2-uB)}.
\end{aligned}
 \label{eq:drift_difference}
\end{equation} Near extinction this is an $O(\Omega^{-1})$ linear correction. The pair-count expression is an algebraic continuation of the weak-return family to $\ell=-u$, not a physical continuous concentration law throughout the quadrant: $\phi_{-u}'(0)=-u$, and $\phi_{-u}(B)<0$ on the unattainable interval $0<B<u$. It therefore lies outside the hypotheses of Theorem~\ref{thm:fold}. On the physical count lattice it vanishes at the singleton state $B=u$ and is nonnegative for $B\geq2u$.

Direct maximization on $B>u$---equivalently, algebraic continuation of Eq.~\eqref{eq:leaky_thresholds}---gives the finite-$\Omega$ drift-skeleton fold
\begin{equation}
 B_{{\rm sn},P}^{\rm drift}=1+\Omega^{-1},\qquad
 \Gamma_{{\rm sn},P}^{\rm drift}=\sqrt{2+\Omega^{-1}}.
 \label{eq:pair_drift_fold}
\end{equation}
At $\Omega=16$, the fold is at $B=\PairDriftB$ ($N_B=17$) and $\Gamma=\PairDriftFold$, only $\PairDriftFoldShift$ above the density-Hill fold. Relative to their own drift folds, the interpolated half-success drives exceed onset by $\HillOwnFoldExcess$ and $\PairOwnFoldExcess$, leaving a residual difference $\OwnFoldResidualGap$; the full half-success drives differ by $\LawShiftEtaTwo$. The fold displacement therefore accounts for only a small part of the operational separation. Together with the factor $\SingleCRatio$ difference in the one-catalyst committors and the isolated-arrival calculation below, this supports attributing most of the remaining separation to low-count eligibility and first-passage dynamics rather than to the drift-fold shift alone.

The concentration limit accordingly gives no relative-error guarantee for initiation from a few molecules \cite{KimJosicBennett2014}. Moreover, the density-Hill singleton event is incompatible with the explicit hard-dimer mechanism in Appendix~\ref{app:massaction}, for which two $B$ molecules are required before the return reaction can fire. The pair-count law respects that eligibility constraint, although it is still not derived by stochastic elimination of the binding intermediates. At $\Omega=16$, the illustrative mean seed $S=0.20$ corresponds to only $3.2$ molecule-equivalents before count preparation.

\subsection{From basins to hitting probabilities}

Every rate vanishes at exact zero in the source-free process. Noise cannot initiate a reaction there without material input. For a finite seed, define extinction $\mathcal A_0=\{(0,0)\}$ and an organized target $\Aorg$, chosen as an ellipse of relative radius $0.20$ about $(C_+,B_+)$. At $\Gamma=2$, $\Omega=16$, its semiaxes are $\TargetSemiC$ and $\TargetSemiB$ molecule counts in the $C$ and $B$ directions. Throughout this section, ``establishment" means first entrance into this target before the stated failure event or deadline; residence and long-term persistence are separate observables. The probability of reaching it before extinction is the committor
\begin{equation}
 q_\Omega(\bm n)=\Prob_{\bm n}(\tau_+<\tau_0),\qquad
 \mathcal L_\Omega q_\Omega=0,
 \label{eq:committor}
\end{equation}
with $q_\Omega=0$ on $\mathcal A_0$ and $q_\Omega=1$ on $\Aorg$ \cite{EVandenEijnden2006,Metzner2009}. We solve the backward equation on a truncated count grid; numerical and target checks are reported in Appendix~\ref{app:numerics}.

Figures~\ref{fig:stochastic}(a,b) show the transition sharpening toward the deterministic separatrix as $\Omega$ increases. The mean Euclidean separation, measured in the normalized $(C,B)$ coordinates, between the separatrix and the $q_\Omega=1/2$ contour [Fig.~\ref{fig:stochastic}(c)] decreases from $\DistanceSmall$ at $\Omega=8$ to $\DistanceLarge$ at $\Omega=64$. A log--log fit over the computed range gives $\Omega^{-\ContourExponent\pm\ContourExponentSE}$, consistent with approximately inverse-system-size convergence but not claimed as an asymptotic exponent. This is the expected weak-noise correspondence away from the basin boundary \cite{FreidlinWentzell2012,Dykman1994}. For a singleton seed $(0,1)$, the pair-count model instead gives exactly zero success, because only loss can occur; the density-Hill model gives $\SingleBHill$ at $\Gamma=2$, $\Omega=16$.

A distribution of prepared counts gives
\begin{equation}
 P_{\rm est}^{\rm stoch}(\Gamma,\Omega;\pi_0)
 =\sum_{\bm n}\pi_0^{(\Omega)}(\bm n)q_\Omega(\bm n).
 \label{eq:Pstoch}
\end{equation}
The baseline rounds $\Omega C_0,\Omega B_0$ to nearest integers. Figure~\ref{fig:stochastic}(d) uses the balanced seeds of Sec.~\ref{sec:basin}. Alternative count-preparation rules are retained in the replication archive; specifying how concentrations become counts is part of the preparation protocol.

\subsection{Sources, deadlines, and operational onset}
\label{sec:finite_sensitivity}

To initiate from exact zero, add catalyst arrivals $\varnothing\xrightarrow{\eta}C$ at total rate $\eta$. The reference $\eta=0.02$ gives a mean interval of $50$ loss-time units while empty. At fixed concentration input $s_C$, the corresponding total rate is $\eta=\Omega s_C$.

Before treating the jump process, augment the common macroscopic drift by the concentration source corresponding to the baseline finite-size protocol, $s_C=\eta/\Omega=\SourceConcentration$:
\begin{subequations}
\label{eq:source_ode}
\begin{align}
 \dot C&=\Gamma\frac{B^2}{1+B^2}-C+s_C,\\
 \dot B&=\Gamma C-B.
\end{align}
\end{subequations}
At equilibrium, $C=B/\Gamma$ and
\begin{equation}
 B^3-(\Gamma^2+\Gamma s_C)B^2+B-\Gamma s_C=0.
 \label{eq:source_cubic}
\end{equation}
The fold conditions admit the parameterization
\begin{equation}
 \Gamma s_C=\frac{B(1-B^2)}{2},\qquad
 \Gamma^2=\frac{(1+B^2)^2}{2B}.
 \label{eq:source_fold_param}
\end{equation}
For $s_C=\SourceConcentration$, the two folds are at $\Gamma=\SourceFoldLow$ and $\SourceFoldHigh$. At $\Gamma=2$, the stable low state $(C,B)=(\SourceCLow,\SourceBLow)$ coexists with the saddle $(\SourceCSaddle,\SourceBSaddle)$ and stable organized state $(\SourceCHigh,\SourceBHigh)$. Deterministic dynamics starting from zero approaches the low source-maintained state rather than the organized branch. The finite-copy transitions computed below are therefore fluctuation- and discrete-arrival-enabled relative to this macroscopic constant-source comparison, not consequences of a deterministic loss of stability of the source-maintained background.

\begin{figure*}[t]
\includegraphics{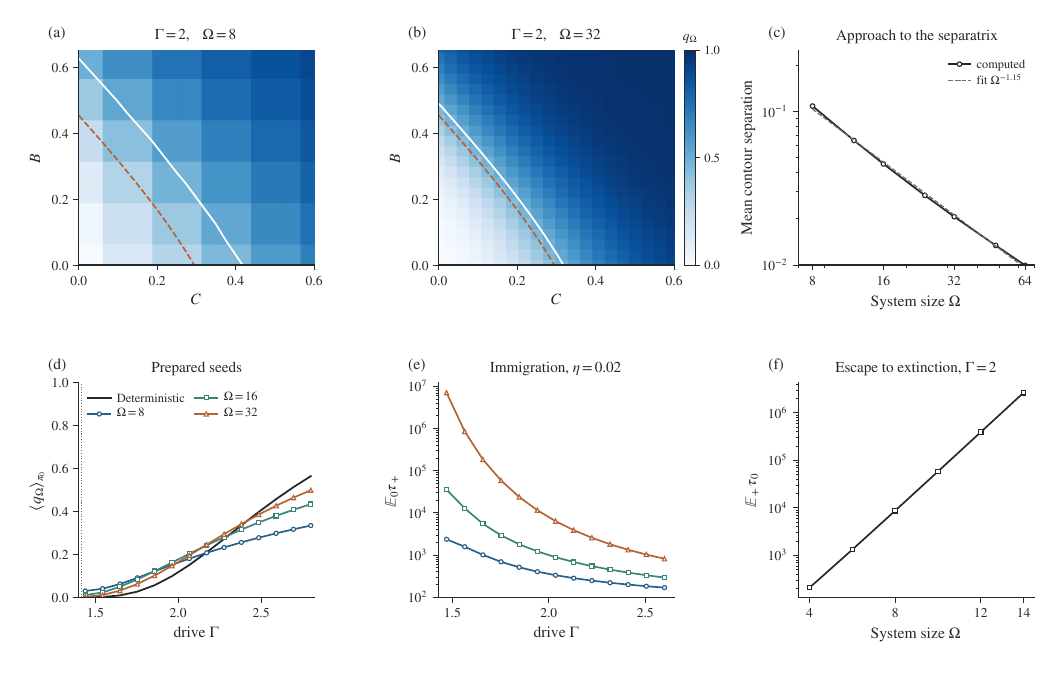}
\caption{Stochastic establishment for the normalized density-Hill process. (a,b) Finite-grid committors at $\Gamma=2$ and $\Omega=8,32$; each colored cell represents one molecular-count state (grids $33\times50$ and $102\times171$, respectively). White contours are $q_\Omega=1/2$, and dashed orange curves are the deterministic separatrix. (c) Mean Euclidean separation between these contours and the separatrix in the normalized $(C,B)$ coordinates at $\Gamma=2$, with a finite-range power-law fit. (d) Balanced-seed averages, using eight scrambled Sobol replicates of $2^{15}$ points; bands measure quadrature uncertainty. (e) Mean entrance time from zero with total catalyst immigration $\eta=0.02$; colors and markers denote the same system sizes as in (d). (f) Source-free mean escape time from the organized equilibrium to extinction at $\Gamma=2$. Escape is shown only through $\Omega=14$ because both the mean time and the count-domain requirements grow rapidly with system size. Panels (e) and (f) measure distinct events. Appendix~\ref{app:numerics} gives numerical checks.}
\label{fig:stochastic}
\end{figure*}

In the dilute-arrival limit, isolated attempts succeed with probability $q_\Omega(1,0)$, giving a leading waiting time $[\eta q_\Omega(1,0)]^{-1}$ when waiting dominates each excursion. Thinning the Poisson arrivals gives the finite-horizon approximation
\begin{equation}
 P_{\rm iso}(T)\simeq1-\exp[-\eta q_\Omega(1,0)T].
 \label{eq:isolated_attempt}
\end{equation}
This connects the calculation to low-immigration colonization \cite{Beer2015}. Our numerical solution includes overlapping arrivals and full excursion times:
\begin{equation}
 \mathcal L_{\Omega,\eta}m_\Omega=-1,\qquad m_\Omega|_{\Aorg}=0,
 \label{eq:mfpt}
\end{equation}
where $m_\Omega(\bm n)=\E_{\bm n}\tau_+$. Increasing drive shortens these times in Fig.~\ref{fig:stochastic}(e). Their increase with $\Omega$ is conditional on fixed total immigration; the concentration input then decreases as $\eta/\Omega$. Figure~\ref{fig:stochastic}(f) shows a separate quantity, source-free escape to extinction, whose mean time grows with system size.

Finite-time establishment is $\Prob(\tau_+\leq T)$ with immigration, or $\Prob(\tau_+<\min\{\tau_0,T\})$ for a source-free seed. If $A$ is the subgenerator outside the absorbing targets and $h$ the rates into $\Aorg$, its success vector obeys
\begin{equation}
 \partial_T u=Au+h,\qquad u(0)=0.
 \label{eq:backward_finite}
\end{equation}
Zero is an absorbing failure target without immigration and a transient state with it. Direct integration of this equation gives the passage-time distribution; a mean waiting time alone does not determine it.

\begin{figure*}[t]
\includegraphics{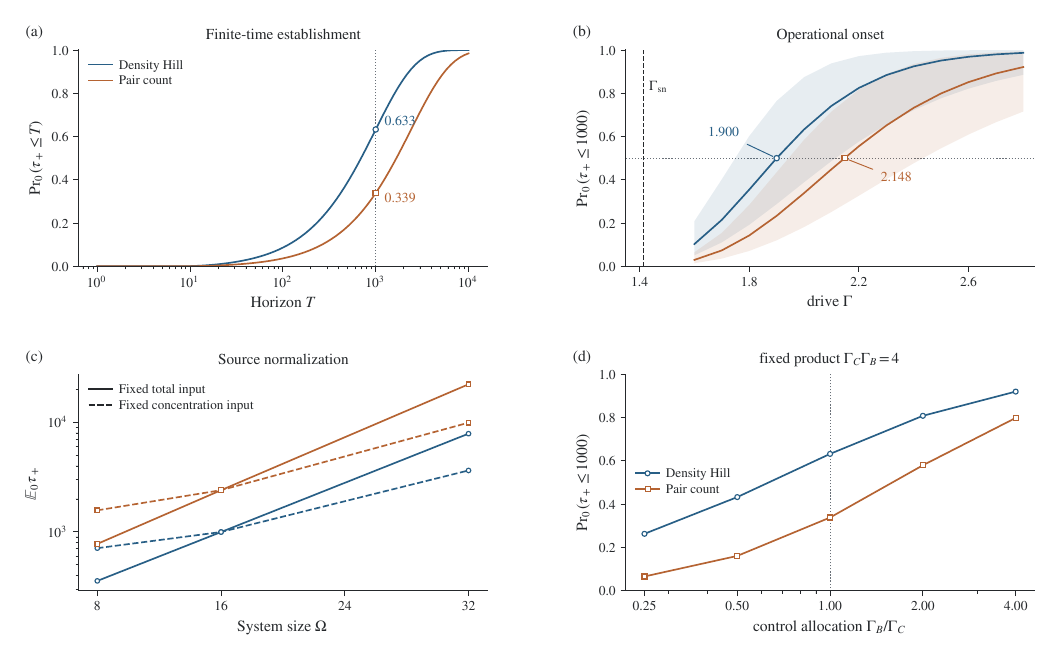}
\caption{Assumptions controlling operational establishment. Blue and orange denote density-Hill and pair-count kinetics throughout. (a) Finite-time entrance from zero at $\Gamma=2$, $\Omega=16$, $\eta=0.02$; markers identify the reported probabilities at $T=1000$. (b) Success at $T=1000$ versus drive: lines use $\eta=0.02$, shading spans $0.01\leq\eta\leq0.04$, and markers locate the interpolated half-success crossings. The vertical dashed line is the common \emph{source-free} fold; shading represents parameter sensitivity. (c) Mean entrance time at fixed total or fixed concentration input (solid and dashed, respectively), matched at $\Omega=16$. (d) Finite-time success at fixed $\Gamma_C\Gamma_B=4$ as the control is redistributed between the cooperative return and feed-forward legs. Panels (a,c) use $\Gamma=2$; all finite-time panels use $\Omega=16$. Quantities are in the normalized units of Sec.~\ref{sec:canonical}.}
\label{fig:sensitivity}
\end{figure*}

At $\Gamma=2$ and $\Omega=16$, the one-catalyst committors are $q_{\Omega,H}(1,0)=\SingleCHill$ and $q_{\Omega,P}(1,0)=\SingleCPair$, a factor $\SingleCRatio$ apart. Equation~\eqref{eq:isolated_attempt} therefore predicts $\IsolatedHill$ and $\IsolatedPair$ at $\eta=0.02$ and $T=1000$. The full calculations give $\FiniteHill$ and $\FinitePair$ [Fig.~\ref{fig:sensitivity}(a)], only $\IsolatedGapHill$ and $\IsolatedGapPair$ higher in absolute probability. At these parameters, the leading model dependence is localized to the eligibility of the first few productive events. The remaining difference reflects departures from isolated attempts, including overlapping arrivals and finite excursion durations. The total $0.294$ separation is far larger than the tested target-radius effect ($1.25\times10^{-3}$) and numerical changes (below $10^{-6}$).

The effect shifts an operational boundary. Table~\ref{tab:onset} reports sampled drive values crossing probability $1/2$ at a fixed deadline. At $\eta=0.02$, interpolation gives $\widehat{\Gamma}_{50}=\GammaFiftyHill$ and $\GammaFiftyPair$, although both source-free deterministic limits have fold $\sqrt2$. Changing the microscopic law shifts the crossing by $\LawShiftEtaTwo$. Halving or doubling $\eta$ shifts it by $\HillEtaHalfShift$ or $\HillEtaDoubleShift$ for the density-Hill law and by $\PairEtaHalfShift$ or $\PairEtaDoubleShift$ for the pair-count law. Source protocol and microscopic eligibility therefore have comparable leverage. For the baseline concentration source, the low deterministic background remains stable until the upper fold $\Gamma=\SourceFoldHigh$; the reported half-success crossings are therefore not deterministic branch-loss thresholds. Figure~\ref{fig:sensitivity}(c) distinguishes fixed total from fixed concentration input; panel (d) tests how a fixed product $\Gamma_C\Gamma_B=4$ is allocated between the two productive legs.

\begin{table}[!htbp]
\caption{Finite-time onset from zero, $\Omega=16$, $T=1000$. Brackets contain the first sampled crossing of $P_{\rm est}=1/2$; $\widehat{\Gamma}_{50}$ is a linear interpolation, not a certified continuum threshold. Both laws share the source-free fold $\sqrt2$.}
\label{tab:onset}
\begin{tabular}{lccc}
\toprule
Return law & $\eta$ & Drive bracket & $\widehat{\Gamma}_{50}$ \\
\midrule
Density Hill & 0.01 & $(2.1,2.2]$ & 2.115 \\
Density Hill & 0.02 & $(1.8,1.9]$ & 1.900 \\
Density Hill & 0.04 & $(1.7,1.8]$ & 1.749 \\
Pair count & 0.01 & $(2.4,2.5]$ & 2.434 \\
Pair count & 0.02 & $(2.1,2.2]$ & 2.148 \\
Pair count & 0.04 & $(1.9,2.0]$ & 1.943 \\
\bottomrule
\end{tabular}

\end{table}

\section{Robustness of finite-amplitude onset}
\label{sec:general_theory}

\subsection{Control path and losses}\label{sec:control_path}

The canonical fold depends on the product of the two productive controls. With independent multipliers $\Gamma_C,\Gamma_B$ its condition is
\begin{equation}
 ab\Gamma_C\Gamma_B=2Kd_Cd_B.
 \label{eq:separate_drive}
\end{equation}
The square root in Eq.~\eqref{eq:Gsn_canonical} follows along $\Gamma_C=\Gamma_B=\Gamma$, with exact sensitivity
\begin{equation}
 \dd\ln \Gamma_{\rm sn}=\tfrac12(\dd\ln K+\dd\ln d_C+\dd\ln d_B
 -\dd\ln a-\dd\ln b).
 \label{eq:fold_sensitivity}
\end{equation}
A factor-two change in one input shifts the fold by $\sqrt2$ or its inverse. The stochastic source protocol is also path dependent because arrivals inject $C$. At fixed $\Gamma_C\Gamma_B=4$, changing $(\Gamma_C,\Gamma_B)$ from $(4,1)$ to $(1,4)$ raises the $T=1000$ success probability from $\ControlHillReturnHeavy$ to $\ControlHillFeedforwardHeavy$ for the density-Hill law and from $\ControlPairReturnHeavy$ to $\ControlPairFeedforwardHeavy$ for the pair-count law [Fig.~\ref{fig:sensitivity}(d)]. The diagonal choice $\Gamma_C=\Gamma_B$ is therefore a specified control path, not a chemical invariant. For the corresponding constant-source system at the baseline $s_C$, annihilation of the low state with the saddle at fixed $\Gamma_C\Gamma_B=4$ requires $\Gamma_Bs_C=\ControlSourceCriticalR$, or $\Gamma_B=\ControlSourceCriticalGammaB$; since Fig.~\ref{fig:sensitivity}(d) uses $\Gamma_B\leq4$, the low deterministic background remains stable throughout that panel. Basin geometry can change even at a fixed fold. For $d_Cd_B=1$ and $\Gamma=2$, the pure-catalyst intercept is $\LossCSymmetric$ at the symmetric point $d_C/d_B=1$ and takes the same value, $\LossCQuarter$, at both sampled extremes $d_C/d_B=0.25$ and $4$. This equality is exact. Equation~\eqref{eq:second_order} depends on the losses only through $d_C+d_B$ and $d_Cd_B$, and a pure-catalyst seed supplies $B(0)=0$, $\dot B(0)=b\Gamma C_0$, independent of the losses; the $C$-axis intercept is therefore invariant under $d_C\leftrightarrow d_B$. A pure-$B$ seed instead has $\dot B(0)=-d_BB_0$, so no such symmetry holds, and over the same range the pure-$B$ intercept decreases from $\LossBQuarter$ to $\LossBFour$. The fold therefore does not summarize preparation sensitivity.

\subsection{Cooperative order and feed-forward amplification}
\label{sec:orders}

The proof of Theorem~\ref{thm:fold} is given in Appendix~\ref{app:chain}. For the Hill family
\begin{equation}
 \phi_p(x)=\frac{x^p}{K^p+x^p},\qquad p>1,
 \label{eq:hillp}
\end{equation}
Theorem~\ref{thm:fold} gives
\begin{equation}
 x_{\rm sn}=K(p-1)^{1/p},\qquad
 \chi_{\max}=\frac{(p-1)^{(p-1)/p}}{pK},
 \label{eq:hillmax}
\end{equation}
and
\begin{equation}
 \Gamma_{\rm sn}=\left[
 \frac{pK D}{A(p-1)^{(p-1)/p}}
 \right]^{1/n}.
 \label{eq:hillGsn}
\end{equation}
If $\phi(x)\sim cx^p$ near zero, the lower finite-amplitude branch obeys
\begin{equation}
 x_{n,-}^*\sim
 \left(\frac{D}{Ac}\right)^{1/(p-1)}
 \Gamma^{-n/(p-1)}.
 \label{eq:lower_scaling}
\end{equation}
Thus increasing drive moves the lower branch toward extinction while leaving $R$ and $\lambda_{\rm kin}$ unchanged. The theorem does not determine the higher-dimensional basin geometry.

\begin{figure*}[t]
\includegraphics{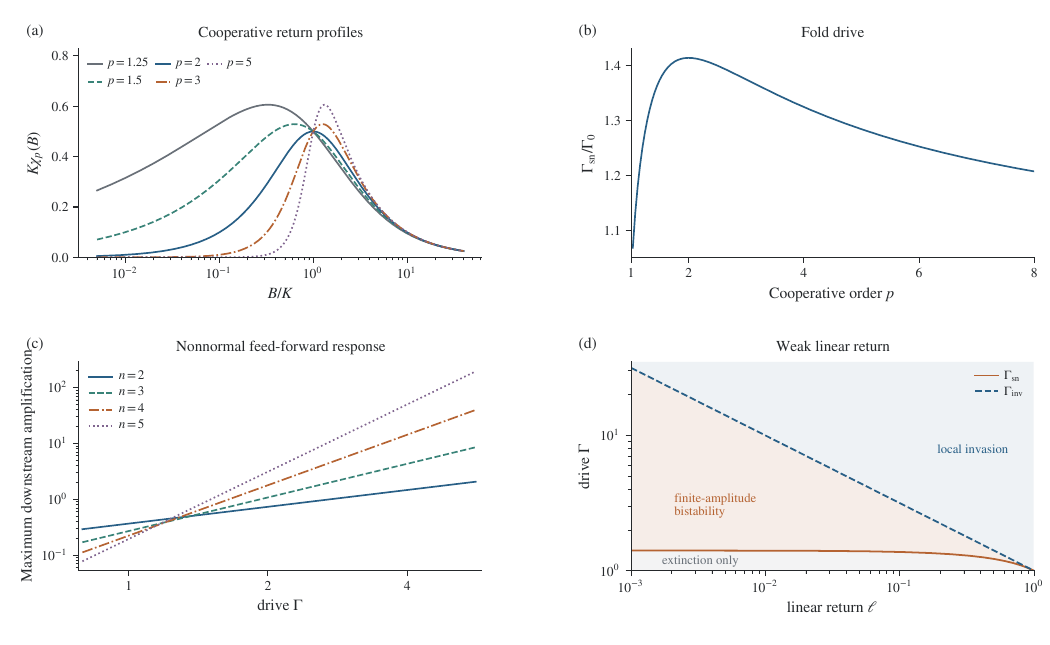}
\caption{Generalization beyond the Hill-2 example. (a) Return profile $\chi_p(B)=\phi_p(B)/B$ for several cooperative orders; a fold requires a positive maximum. (b) Dimensionless fold drive for the two-stage Hill family, $\Gamma_{\rm sn}/\Gamma_0=[p/(p-1)^{(p-1)/p}]^{1/2}$ with $\Gamma_0=(Kd_Cd_B/ab)^{1/2}$. (c) Maximum linear downstream response of equal-loss feed-forward chains from Eq.~\eqref{eq:nstage_max}. All eigenvalues remain negative while the transient response grows with drive for fixed chain length. Increasing depth can amplify or attenuate the response, depending on $a_f\Gamma/d$; the plot uses $a_f=d=1$. (d) Phase structure after adding a linear return of weight $\ell$: below $\Gamma_{\rm sn}$ only extinction exists, between $\Gamma_{\rm sn}$ and $\Gamma_{\rm inv}$ extinction coexists with a finite-amplitude organized branch, and above $\Gamma_{\rm inv}$ extinction is locally unstable.}
\label{fig:general}
\end{figure*}

Figure~\ref{fig:general}(a,b) displays the Hill-family return profile and fold from Eqs.~\eqref{eq:hillmax}--\eqref{eq:hillGsn}.

Feed-forward depth creates an additional effect. The linearized $n$-stage chain is lower triangular, but it is generally nonnormal. With equal losses $d_i=d$ and equal feed-forward rates $a_i=a_f$ for $i\geq2$, the linearized response to a seed $x_{1,0}$ in the first species is
\begin{equation}
 \frac{x_n(t)}{x_{1,0}}=(a_f\Gamma)^{n-1}\frac{t^{n-1}}{(n-1)!}e^{-dt},
 \label{eq:nstage_transient}
\end{equation}
which reaches
\begin{equation}
 \max_t\frac{x_n(t)}{x_{1,0}}=
 \left(\frac{a_f\Gamma}{d}\right)^{n-1}
 \frac{(n-1)^{n-1}e^{-(n-1)}}{(n-1)!}.
 \label{eq:nstage_max}
\end{equation}
The eigenvalues remain $-d$, and $R=0$, while transient amplification increases as $\Gamma^{n-1}$. For a seed already in the organized basin, this feed-forward response can bring the return reaction into its nonlinear regime. An autonomous deterministic trajectory cannot cross an invariant basin boundary. Local asymptotic stability therefore does not imply weak finite-time response.

\subsection{Weak linear return}
\label{sec:leakage}

To perturb the exact-zero return derivative while retaining saturation, use
\begin{equation}
 \phi_\ell(B)=\frac{\ell B+B^2}{1+\ell B+B^2},\qquad0\leq\ell<1,
 \label{eq:leaky_return}
\end{equation}
in $\dot C=\Gamma\phi_\ell(B)-C$, $\dot B=\Gamma C-B$. The dimensionless weight $\ell$ defines a structural sensitivity family; $\ell=0$ recovers the original model. Extinction now has
\begin{equation}
 R=\Gamma\sqrt\ell,\qquad\lambda_{\rm kin}=-1+\Gamma\sqrt\ell.
 \label{eq:leaky_local}
\end{equation}
Positive equilibria solve
\begin{equation}
 B^2+(\ell-\Gamma^2)B+1-\ell \Gamma^2=0.
 \label{eq:leaky_equilibrium}
\end{equation}
The return profile $(\ell+B)/(1+\ell B+B^2)$ is maximal at $B=1-\ell$, giving
\begin{equation}
 \Gamma_{\rm sn}(\ell)=\sqrt{2-\ell},\qquad
 \Gamma_{\rm inv}(\ell)=\ell^{-1/2}\quad(\ell>0).
 \label{eq:leaky_thresholds}
\end{equation}
For $0<\ell<1$, the interval between these thresholds has stable extinction, a saddle, and a stable productive state. At $\ell=0.01$, for example, the fold is $\sqrt{1.99}$ and invasion occurs at $10$. A finite-amplitude regime thus survives a weak linear return, while the fixed negative exponent at all drive values is specific to $\ell=0$. The interval closes as $\ell\uparrow1$. These formulas are checked by independent roots and Jacobian eigenvalues; the stochastic figures retain $\ell=0$.

The maintenance mechanism also extends to a feed-forward chain closed by a cooperative return $\phi$. Theorem~\ref{thm:fold} gives the exact conditions: a single nondegenerate maximum of $\phi(x)/x$ produces the fold, and logarithmic return slope $x\phi'(x)/\phi(x)<1$ gives a stable positive branch. Figure~\ref{fig:general}(d) shows the finite interval between branch creation and local invasion when a weak linear return is added. The theorem delimits the model class; higher-dimensional basins and molecular establishment probabilities still require separate calculations.

\section{Discussion and conclusions}
\label{sec:discussion}\label{sec:thresholds}

The analysis separates three thresholds that coincide only in special models. The \emph{existence threshold} $\Gamma_{\rm sn}$ creates the saddle and organized branch. Deterministic \emph{accessibility} is geometrical: a specified state lies in $\Borg(\Gamma)$, or an uncertain preparation has the basin weight in Eq.~\eqref{eq:Pdet}; a finite horizon further restricts this to the first-entrance measure in Eq.~\eqref{eq:Pdet_finite}, evaluated in the inset of Fig.~\ref{fig:detprob}(a). Finite-time stochastic \emph{establishment} is probabilistic and depends additionally on $\Omega$, the molecular generator, the seed or source protocol, the target, and the observation window. Equal-amount seeds can have opposite deterministic outcomes, and deterministically equivalent jump laws can differ strongly at low copy number.

\subsection{Extension of the kinetic component of Gate 7}
\label{sec:gate7}

Gate~7 in our earlier framework \cite{Turyshev2026} is a conjunction of structural and stoichiometric closure, thermodynamic feasibility, and kinetic accessibility. This paper addresses only the kinetic leg: it neither evaluates the RAF-percolation proxy $p_{\rm cat}>p_c$ nor establishes a complete dissipative flux with $\sum_i v_i\Delta G_i<0$ \cite{HornJackson1972,RaoEsposito2016}. The local threshold conventions are the same in the two papers, namely $\lambda_{\rm kin}=\al(A_0)$ and $R=\rho(J_{\rm prod}\Lambda^{-1})$. The canonical model has $\lambda_{\rm kin}=-1$ and $R=0$ for every $\Gamma$, yet a stable organized branch is reachable from finite seeds for $\Gamma>\sqrt2$. It is therefore a counterexample to treating positive local invasion as a mechanism-independent necessary condition for kinetic establishment, not a disagreement about definitions and not a demonstration that the other Gate~7 requirements are met.

Conditional on an independently established structurally closed, dissipative maintenance mechanism, retain $\lambda_{\rm kin}$ and $R$ as local-invasion diagnostics and define an operational kinetic margin
\begin{equation}
 m_{7,\mathrm{est}}
 =P_{\rm est}(T;\Gamma,\Omega,\mathcal P,\Aorg)-p_*,
 \label{eq:gate7_est_margin}
\end{equation}
where $\mathcal P$ specifies a seed, immigration process, or other source protocol. Passage requires $m_{7,\mathrm{est}}\geq0$ for the stated generator, target, and deadline. If monotonicity in $\Gamma$ is established for a fixed protocol and target, one may define
\begin{equation}
 \Gamma_{\rm est}
 =\inf\{\Gamma>\Gamma_{\rm sn}:m_{7,\mathrm{est}}\geq0\}.
 \label{eq:Gest}
\end{equation}
The relative ellipse used here moves outward with the organized equilibrium, so increasing birth rates does not by itself prove monotonicity of this moving-target probability. Table~\ref{tab:onset} therefore reports sampled, interpolated crossings $\widehat{\Gamma}_{50}$ rather than certified continuum thresholds. For the birth--death generators used here, a fixed upward-closed target restores monotonicity in the common drive by the standard order-preserving coupling.

With a specified initial distribution, Eq.~\eqref{eq:Pdet} is the Gate~7 analogue of the basin fraction already used for multistability at Gate~12 of Ref.~\cite{Turyshev2026}. The first-passage calculation adds molecular eligibility and a deadline. This dependence is also consistent with the earlier paper's classification of Gate~7 margins as prior dominated: remote observations alone do not determine $\pi_0$, $\Omega$, the microscopic generator, or the source history. Confinement can alter losses, molecular counts, the fold, and basin geometry; multiplying existing feed-forward coefficients cannot create a missing linear cycle, whereas a new linear return changes the productive graph, as in Sec.~\ref{sec:leakage}.

\subsection{Physical interpretation and limitations}

The deterministic result is not tied to the Hill-2 law. Superlinear onset $\phi(x)\sim cx^p$ with $p>1$ removes the return edge from the productive linearization, while a finite-amplitude fold depends on the global shape of $\phi(x)/x$. Saturation is one sufficient way to make that return profile fall at large abundance. The weak-linear-return family shows that bistability persists when the local graph is perturbed: there is then a finite interval between branch creation and local invasion.

The principal physical limitation is microscopic. The density-Hill and pair-count processes are effective jump models, and their difference quantifies coarse-graining sensitivity rather than uncertainty within a single chemical mechanism. In particular, density-Hill kinetics permits catalyst production from a single $B$ molecule and therefore cannot be the finite-copy realization of the hard-dimer mechanism in Appendix~\ref{app:massaction}; pair-count kinetics has the correct two-copy eligibility but is still not a stochastic elimination of that mechanism. The deterministic mass-action construction requires fast equilibration and low sequestration; at the baseline stochastic size its site inventory is $\Omega\varepsilon=\OmegaEpsilon\ll1$. A predictive molecular model therefore requires an integer-site realization and volume-consistent elementary rates. The system is well mixed and chemostatted. The Lyapunov potential in Appendix~\ref{app:proofs} is dynamical, not an activation free energy, so temperature dependence must enter through the kinetic and equilibrium parameters rather than through that potential.

The experimentally accessible signatures are protocol specific: a finite seeded down-ramp collapse despite no deterministic up-ramp from exact extinction, composition-dependent success at fixed total seed, a committor transition approaching the deterministic separatrix with increasing system size, and different onset statistics for molecular mechanisms sharing the same concentration limit. For planetary applications, the process-level probability in Eq.~\eqref{eq:gate7_est_margin} would require compartment sizes, rates, preparation statistics, and environmental durations; it cannot be identified directly with a planet's probability of hosting life or used without recalibration in a target score \cite{Catling2018}.

In summary, local viability and global establishment answer different questions. When productive closure first appears nonlinearly, extinction may remain strictly stable with $R=0$ while an organized branch exists at finite amplitude. Branch existence is then characterized by nonlinear branch structure, deterministic accessibility by basin geometry, and molecular establishment by committor and first-passage probabilities under a specified protocol.

\section*{Acknowledgments}
The research was carried out in part at the Jet Propulsion Laboratory, California Institute of Technology, under a contract with the National Aeronautics and Space Administration. 
\textcopyright\ 2026. California Institute of Technology. Government sponsorship acknowledged.

\appendix

\section{Proof of Proposition~\ref{prop:acyclic} and canonical convergence}
\label{app:proofs}

\subsection{Proof of Proposition~\ref{prop:acyclic}}

A topological ordering permutes an acyclic nonnegative matrix into strictly triangular form. Right multiplication by the positive diagonal matrix $\Lambda^{-1}$ preserves strict triangularity, so $K_0=J_{\rm prod}\Lambda^{-1}$ is nilpotent and $R=0$. In the same ordering, $A_0=J_{\rm prod}-\Lambda$ is triangular with diagonal entries $-d_i$, proving Eq.~\eqref{eq:acyclic_result}.

\subsection{Scaling and global convergence}

To identify the independent deterministic control parameters, introduce
\begin{align}
 \tau&=d_Bt, & y&=\frac{B}{K}, &
 x&=\frac{b\Gamma C}{d_BK},\nonumber\\
 \delta&=\frac{d_C}{d_B}, &
 \mu&=\frac{ab\Gamma^2}{Kd_Cd_B}.
 \label{eq:nondim}
\end{align}
This gives
\begin{equation}
 x'=\delta\left(\mu\frac{y^2}{1+y^2}-x\right),
 \qquad y'=x-y.
 \label{eq:nondim_system}
\end{equation}
The reference coefficients give $\delta=1$ and $\mu=\Gamma^2$. The transformed coordinate $x$ depends on the drive, so the seed distributions are specified in the fixed $(C,B)$ coordinates of Sec.~\ref{sec:canonical}.

The physical quadrant is forward invariant: at $C=0$, $\dot C\geq0$, and at $B=0$, $\dot B=b\Gamma C\geq0$. Eliminating $C$ gives the damped second-order equation
\begin{equation}
 \ddot B+(d_C+d_B)\dot B+d_Cd_BB
 -ab\Gamma^2\frac{B^2}{K^2+B^2}=0.
 \label{eq:second_order}
\end{equation}
Define
\begin{align}
 V(B)&=\frac{d_Cd_B}{2}B^2
 -ab\Gamma^2\left[B-K\arctan\left(\frac{B}{K}\right)\right],\label{eq:potential}\\
 E(B,\dot B)&=\frac12\dot B^2+V(B).
 \label{eq:energy}
\end{align}
Then
\begin{equation}
 \dot E=-(d_C+d_B)\dot B^2\leq0.
 \label{eq:energy_decay}
\end{equation}
Because $V(B)\to+\infty$ as $B\to\infty$, trajectories are bounded. LaSalle's invariance principle implies convergence to equilibria. Independently, the planar divergence is the negative constant $-(d_C+d_B)$, so the Bendixson criterion excludes periodic orbits. For this planar model, these facts give convergence to one of the finitely many equilibria and exclude additional attractors. They justify the basin classification independently of a finite integration cutoff. The function $V$ is a dynamical Lyapunov potential, not an activation free energy: it does not supply a stochastic barrier or a temperature law.

\section{Proof and consequences of Theorem~\ref{thm:fold}}
\label{app:chain}

\subsection{Equilibria and stability}

For Eq.~\eqref{eq:nchain}, the productive Jacobian at extinction contains only the edges $x_{i-1}\to x_i$ for $i\geq2$, because $\phi'(0)=0$. Proposition~\ref{prop:acyclic} gives the extinction result.

At a positive equilibrium, recurrence through the feed-forward chain yields
\begin{equation}
 x_i^*=\frac{a_i\Gamma}{d_i}x_{i-1}^*,\qquad
 x_n^*=\frac{\Gamma^{n-1}\prod_{i=2}^{n}a_i}{\prod_{i=2}^{n}d_i}x_1^*.
 \label{eq:chain_recurrence}
\end{equation}
Substitution into the $x_1$ equation gives Eq.~\eqref{eq:qintersection}. Since $D/(A\Gamma^n)$ decreases monotonically from infinity to zero and $\chi$ has one nondegenerate maximum, the intersection count is zero, one, or two as stated. At equality with $\chi_{\max}$, transversality in $\Gamma$ and $\chi''(x_{\rm sn})<0$ give a generic saddle-node \cite{Kuznetsov2004,Strogatz2015}.

At a positive equilibrium, write the Jacobian as $M=F_*-\Lambda$. The nonnegative matrix $F_*$ contains a single directed cycle with edge weights $a_i\Gamma$ for $i=2,\ldots,n$ and $a_1\Gamma\phi'(x_n^*)$ on the return edge. Hence
\begin{equation}
 \rho(F_*\Lambda^{-1})^n
 =\frac{A\Gamma^n\phi'(x_n^*)}{D}.
 \label{eq:local_cycle_gain}
\end{equation}
Using Eq.~\eqref{eq:qintersection}, the right-hand side is $\nu_\phi(x_n^*)$. Since $F_*\geq0$, $M=F_*-\Lambda$ is Metzler, so its spectral abscissa is a real eigenvalue. The standard Metzler-matrix threshold criterion gives $M$ Hurwitz precisely when $\rho(F_*\Lambda^{-1})<1$, proving the logarithmic-slope test. Consequently, the first loss of stability along this branch occurs through a zero real eigenvalue at $\nu_\phi=1$; a Hopf bifurcation cannot be the first local instability. Finally,
\begin{equation}
 \chi'(x)=\frac{\phi(x)}{x^2}\left[\frac{x\phi'(x)}{\phi(x)}-1\right]
 \label{eq:qprime_elasticity}
\end{equation}
relates the branch slopes to stability. At the nondegenerate fold one simple eigenvalue crosses zero; the lower branch therefore has one unstable direction near onset, and in the planar case it is a saddle throughout.

\section{Full mass-action realization}
\label{app:massaction}

A mass-action realization of the cooperative term is
\begin{align}
 2B &\underset{k_d^-}{\stackrel{k_d^+}{\rightleftharpoons}} D,
 &D+S&\underset{k_s^-}{\stackrel{k_s^+}{\rightleftharpoons}} Q,
 \label{eq:assembly_reactions}\\
 Q+A_C&\longrightarrow Q+C+W_C,
 &C+A_B&\longrightarrow C+B+W_B.
 \label{eq:catalytic_reactions}
\end{align}
With chemostatted substrates $A_C,A_B$, conserved sites $S_T=S+Q$, and rapid equilibration of dimerization and site binding,
\begin{equation}
 D=\frac{B^2}{K_d},\qquad
 Q=S_T\frac{B^2}{K_dK_s+B^2}.
 \label{eq:rapid_eq}
\end{equation}
Setting $K^2=K_dK_s$ and absorbing $S_T$ and chemostat activities into $a\Gamma$ gives Eq.~\eqref{eq:Cdot}. The finite-rate equations below test this deterministic reduction.

\begin{figure*}[t]
\includegraphics{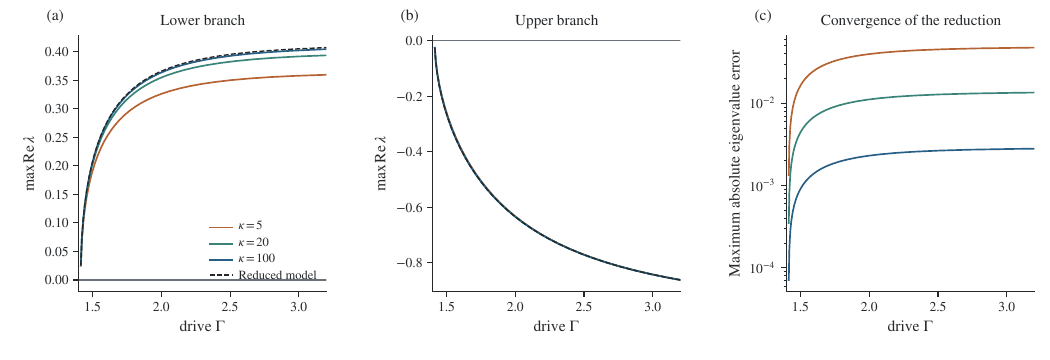}
\caption{Finite-rate mass-action reduction with $L=10^5$ and $\varepsilon=10^{-5}$. (a,b) Dominant Jacobian eigenvalues on the lower and upper equilibrium branches; the branches themselves coincide exactly with the reduced model. (c) At each drive value, the larger of the two absolute differences between the full and reduced dominant eigenvalues. This resolves convergence that is difficult to see where the curves in (b) overlap. Colors denote $\kappa=5,20,100$ in all panels. The lower branch remains unstable and the upper branch stable.}
\label{fig:massaction}
\end{figure*}

For a fast, low-sequestration realization set
\begin{equation}
 K_d=L,\qquad K_s=L^{-1},\qquad S_T=\varepsilon,
 \label{eq:massaction_scales}
\end{equation}
choose fast rates $k_d^+=\kappa$, $k_d^-=\kappa L$, $k_s^+=\kappa L$, and $k_s^-=\kappa$, and take the total catalyst-production rate to be $\Gamma Q/\varepsilon$. The normalized unreduced equations are
\begin{subequations}
\label{eq:full_mass_action}
\begin{align}
 \dot C&=\frac{\Gamma}{\varepsilon}Q-C,\\
 \dot B&=\Gamma C-B-2\kappa B^2+2\kappa L D,\\
 \dot D&=\kappa B^2-\kappa L D
          -\kappa L D(\varepsilon-Q)+\kappa Q,\\
 \dot Q&=\kappa L D(\varepsilon-Q)-\kappa Q.
\end{align}
\end{subequations}
At equilibrium,
\begin{equation}
 D=\frac{B^2}{L},\qquad
 Q=\varepsilon\frac{B^2}{1+B^2},
 \label{eq:massaction_equilibrium}
\end{equation}
so the positive equilibrium branches and fold are exactly those of the reduced model. The total assembly-component inventory is $M=B+2D+2Q$, and the full equations give $\dot M=\Gamma C-B$. On the rapid-equilibrium manifold,
\begin{equation}
 \frac{\dd M}{\dd B}=1+\frac{4B}{L}
 +\frac{4\varepsilon B}{(1+B^2)^2}.
 \label{eq:sequestration}
\end{equation}
Rapid equilibration therefore gives $\dot B=(\Gamma C-B)/(\dd M/\dd B)$; the two-variable model additionally requires small sequestration. For $L=10^5$, $\varepsilon=10^{-5}$ and the plotted branches, the correction to unity is below $4.1\times10^{-4}$. The accompanying turnover is $\Gamma/\varepsilon$. At the baseline stochastic size, $\Omega\varepsilon=\OmegaEpsilon\ll1$, so these deterministic parameters cannot also specify an integer site inventory. Figure~\ref{fig:massaction} verifies branch stability and slow-eigenvalue convergence at finite $\kappa$. The archive supplies the finite-parameter sweep. This is a deterministic reduction check; stochastic elimination requires its own validity conditions \cite{SegelSlemrod1989,KimJosicBennett2014}.

\section{Numerical protocol and reproducibility}
\label{app:numerics}

The archive supplies complete sources, fixed configuration, figure and table data, pinned Python dependencies, and validation records. Calculations use Python 3.12.13, NumPy 2.3.5, SciPy 1.17.0, and Matplotlib 3.10.8. No measured inputs are used. Positive equilibria are analytic for Hill order two and found by bracketed roots otherwise.

Stable manifolds are initialized $10^{-8}$ from the saddle along its stable eigenvector and integrated backward with RK45, relative/absolute tolerances $2\times10^{-11}$/$2\times10^{-13}$ and maximum steps $0.01$ or $0.018$. Direct trajectories use $10^{-10}$/$10^{-12}$ and maximum step $0.05$. Basin labels come from the continued manifold. Seed averages use eight independent scrambled Sobol replicates, with $2^{16}$ samples for asymptotic deterministic accessibility and $2^{15}$ for stochastic averages; scramble seeds begin at $9173$ and $4421$. The deterministic finite-time inset uses eight nested Sobol replicates of $2^{12}$ seeds, a fixed-step fourth-order Runge--Kutta method with $\Delta t=0.05$, and the same relative-radius-$0.20$ target as the stochastic calculation. Halving the step changed no reported cumulative fraction in this sample. Bands are $1.96$ standard errors across replicate means. Count preparation uses nearest integers with ties to even or independent unbiased stochastic rounding. Half-committor crossings are interpolated in both concentration coordinates, and their Euclidean distances to the piecewise-linear separatrix are measured in the normalized $(C,B)$ coordinates.

Committors and mean passage times are solved by sparse direct factorization. The standard count grid extends beyond $(1.55C_+,1.35B_+)$ by eight counts and suppresses outward births. Expanding it to $(1.85C_+,1.65B_+)$ plus twelve counts changes low-abundance committors by less than $2\times10^{-4}$ at $\Gamma=2$. Organized targets are relative-radius-$0.20$ ellipses. Finite-time equations use BDF with sparse Jacobian and relative/absolute tolerances $2\times10^{-8}$/$2\times10^{-10}$.

Finite-time checks use $\Gamma=1.7,2.0,2.4,2.8$, $\Omega=16$, $T=100,1000$, and $\eta=0.04$ at $\Gamma=1.7$ or $0.02$ otherwise. Expanded domains change success probabilities by less than $4.9\times10^{-7}$, tighter integration by less than $8.6\times10^{-7}$, and target radii $0.15$ or $0.25$ by at most $1.25\times10^{-3}$. Source-driven mean times change by less than $3.6\times10^{-7}$ relative on expanded grids; the largest change among six plotted escape times is $7.9\times10^{-4}$. These are measured checks at the stated samples, not uniform bounds.

Independent checks cover backward-equation residuals, basin invariance, the relative and absolute propensity identities, the finite-$\Omega$ drift fold, the source-maintained cubic and its two folds, general-chain stability, and 288 finite-parameter mass-action equilibria. The weak-return calculation uses $\ell=0,0.01,0.1,0.25,0.5,0.9$ and compares analytic branches with bracketed roots and direct eigenvalues; root discrepancies are below $3\times10^{-14}$.


\begin{thebibliography}{33}%
\makeatletter
\providecommand \@ifxundefined [1]{%
 \@ifx{#1\undefined}
}%
\providecommand \@ifnum [1]{%
 \ifnum #1\expandafter \@firstoftwo
 \else \expandafter \@secondoftwo
 \fi
}%
\providecommand \@ifx [1]{%
 \ifx #1\expandafter \@firstoftwo
 \else \expandafter \@secondoftwo
 \fi
}%
\providecommand \natexlab [1]{#1}%
\providecommand \enquote  [1]{``#1''}%
\providecommand \bibnamefont  [1]{#1}%
\providecommand \bibfnamefont [1]{#1}%
\providecommand \citenamefont [1]{#1}%
\providecommand \href@noop [0]{\@secondoftwo}%
\providecommand \href [0]{\begingroup \@sanitize@url \@href}%
\providecommand \@href[1]{\@@startlink{#1}\@@href}%
\providecommand \@@href[1]{\endgroup#1\@@endlink}%
\providecommand \@sanitize@url [0]{\catcode `\\12\catcode `\$12\catcode
  `\&12\catcode `\#12\catcode `\^12\catcode `\_12\catcode `\%12\relax}%
\providecommand \@@startlink[1]{}%
\providecommand \@@endlink[0]{}%
\providecommand \url  [0]{\begingroup\@sanitize@url \@url }%
\providecommand \@url [1]{\endgroup\@href {#1}{\urlprefix }}%
\providecommand \urlprefix  [0]{URL }%
\providecommand \Eprint [0]{\href }%
\providecommand \doibase [0]{https://doi.org/}%
\providecommand \selectlanguage [0]{\@gobble}%
\providecommand \bibinfo  [0]{\@secondoftwo}%
\providecommand \bibfield  [0]{\@secondoftwo}%
\providecommand \translation [1]{[#1]}%
\providecommand \BibitemOpen [0]{}%
\providecommand \bibitemStop [0]{}%
\providecommand \bibitemNoStop [0]{.\EOS\space}%
\providecommand \EOS [0]{\spacefactor3000\relax}%
\providecommand \BibitemShut  [1]{\csname bibitem#1\endcsname}%
\let\auto@bib@innerbib\@empty
\bibitem [{\citenamefont {Turyshev}(2026)}]{Turyshev2026}%
  \BibitemOpen
  \bibfield  {author} {\bibinfo {author} {\bibfnamefont {S.~G.}\ \bibnamefont
  {Turyshev}},\ }\bibfield  {title} {\bibinfo {title} {Quantitative
  nonequilibrium pathway from fundamental physics to the emergence and
  persistence of exoplanetary biospheres},\ }\href
  {https://doi.org/10.1007/s11084-026-09727-7} {\bibfield  {journal} {\bibinfo
  {journal} {Discover Life}\ }\textbf {\bibinfo {volume} {56}},\ \bibinfo
  {pages} {9} (\bibinfo {year} {2026})},\ \Eprint
  {https://arxiv.org/abs/2606.02648} {arXiv:2606.02648 [physics.gen-ph]}
  \BibitemShut {NoStop}%
\bibitem [{\citenamefont {Schl{\"o}gl}(1972)}]{Schlogl1972}%
  \BibitemOpen
  \bibfield  {author} {\bibinfo {author} {\bibfnamefont {F.}~\bibnamefont
  {Schl{\"o}gl}},\ }\bibfield  {title} {\bibinfo {title} {Chemical reaction
  models for non-equilibrium phase transitions},\ }\href
  {https://doi.org/10.1007/BF01379769} {\bibfield  {journal} {\bibinfo
  {journal} {Z. Physik}\ }\textbf {\bibinfo {volume} {253}},\ \bibinfo {pages}
  {147} (\bibinfo {year} {1972})}\BibitemShut {NoStop}%
\bibitem [{\citenamefont {Dennis}(1989)}]{Dennis1989}%
  \BibitemOpen
  \bibfield  {author} {\bibinfo {author} {\bibfnamefont {B.}~\bibnamefont
  {Dennis}},\ }\bibfield  {title} {\bibinfo {title} {Allee effects: Population
  growth, critical density, and the chance of extinction},\ }\href
  {https://doi.org/10.1111/j.1939-7445.1989.tb00119.x} {\bibfield  {journal}
  {\bibinfo  {journal} {Nat. Resour. Model.}\ }\textbf {\bibinfo {volume}
  {3}},\ \bibinfo {pages} {481} (\bibinfo {year} {1989})}\BibitemShut {NoStop}%
\bibitem [{\citenamefont {Ohtsuki}\ and\ \citenamefont
  {Nowak}(2009)}]{OhtsukiNowak2009}%
  \BibitemOpen
  \bibfield  {author} {\bibinfo {author} {\bibfnamefont {H.}~\bibnamefont
  {Ohtsuki}}\ and\ \bibinfo {author} {\bibfnamefont {M.~A.}\ \bibnamefont
  {Nowak}},\ }\bibfield  {title} {\bibinfo {title} {Prelife catalysts and
  replicators},\ }\href {https://doi.org/10.1098/rspb.2009.1136} {\bibfield
  {journal} {\bibinfo  {journal} {Proc. R. Soc. B}\ }\textbf {\bibinfo {volume}
  {276}},\ \bibinfo {pages} {3783} (\bibinfo {year} {2009})}\BibitemShut
  {NoStop}%
\bibitem [{\citenamefont {Giri}\ and\ \citenamefont
  {Jain}(2012)}]{GiriJain2012}%
  \BibitemOpen
  \bibfield  {author} {\bibinfo {author} {\bibfnamefont {V.}~\bibnamefont
  {Giri}}\ and\ \bibinfo {author} {\bibfnamefont {S.}~\bibnamefont {Jain}},\
  }\bibfield  {title} {\bibinfo {title} {The origin of large molecules in
  primordial autocatalytic reaction networks},\ }\href
  {https://doi.org/10.1371/journal.pone.0029546} {\bibfield  {journal}
  {\bibinfo  {journal} {PLOS ONE}\ }\textbf {\bibinfo {volume} {7}},\ \bibinfo
  {pages} {e29546} (\bibinfo {year} {2012})}\BibitemShut {NoStop}%
\bibitem [{\citenamefont {Hordijk}\ and\ \citenamefont
  {Steel}(2004)}]{HordijkSteel2004}%
  \BibitemOpen
  \bibfield  {author} {\bibinfo {author} {\bibfnamefont {W.}~\bibnamefont
  {Hordijk}}\ and\ \bibinfo {author} {\bibfnamefont {M.}~\bibnamefont
  {Steel}},\ }\bibfield  {title} {\bibinfo {title} {Detecting autocatalytic,
  self-sustaining sets in chemical reaction systems},\ }\href
  {https://doi.org/10.1016/j.jtbi.2003.11.020} {\bibfield  {journal} {\bibinfo
  {journal} {J. Theor. Biol.}\ }\textbf {\bibinfo {volume} {227}},\ \bibinfo
  {pages} {451} (\bibinfo {year} {2004})}\BibitemShut {NoStop}%
\bibitem [{\citenamefont {Blokhuis}\ \emph {et~al.}(2020)\citenamefont
  {Blokhuis}, \citenamefont {Lacoste},\ and\ \citenamefont
  {Nghe}}]{Blokhuis2020}%
  \BibitemOpen
  \bibfield  {author} {\bibinfo {author} {\bibfnamefont {A.}~\bibnamefont
  {Blokhuis}}, \bibinfo {author} {\bibfnamefont {D.}~\bibnamefont {Lacoste}},\
  and\ \bibinfo {author} {\bibfnamefont {P.}~\bibnamefont {Nghe}},\ }\bibfield
  {title} {\bibinfo {title} {Universal motifs and the diversity of
  autocatalytic systems},\ }\href {https://doi.org/10.1073/pnas.2013527117}
  {\bibfield  {journal} {\bibinfo  {journal} {Proc. Natl. Acad. Sci. U.S.A.}\
  }\textbf {\bibinfo {volume} {117}},\ \bibinfo {pages} {25230} (\bibinfo
  {year} {2020})}\BibitemShut {NoStop}%
\bibitem [{\citenamefont {Peng}\ \emph {et~al.}(2022)\citenamefont {Peng},
  \citenamefont {Linderoth},\ and\ \citenamefont {Baum}}]{Peng2022}%
  \BibitemOpen
  \bibfield  {author} {\bibinfo {author} {\bibfnamefont {Z.}~\bibnamefont
  {Peng}}, \bibinfo {author} {\bibfnamefont {J.}~\bibnamefont {Linderoth}},\
  and\ \bibinfo {author} {\bibfnamefont {D.~A.}\ \bibnamefont {Baum}},\
  }\bibfield  {title} {\bibinfo {title} {The hierarchical organization of
  autocatalytic reaction networks and its relevance to the origin of life},\
  }\href {https://doi.org/10.1371/journal.pcbi.1010498} {\bibfield  {journal}
  {\bibinfo  {journal} {PLOS Comput. Biol.}\ }\textbf {\bibinfo {volume}
  {18}},\ \bibinfo {pages} {e1010498} (\bibinfo {year} {2022})}\BibitemShut
  {NoStop}%
\bibitem [{\citenamefont {Semenov}\ \emph {et~al.}(2016)\citenamefont
  {Semenov}, \citenamefont {Kraft}, \citenamefont {Ainla}, \citenamefont
  {Zhao}, \citenamefont {Baghbanzadeh}, \citenamefont {Campbell}, \citenamefont
  {Kang}, \citenamefont {Fox},\ and\ \citenamefont {Whitesides}}]{Semenov2016}%
  \BibitemOpen
  \bibfield  {author} {\bibinfo {author} {\bibfnamefont {S.~N.}\ \bibnamefont
  {Semenov}}, \bibinfo {author} {\bibfnamefont {L.~J.}\ \bibnamefont {Kraft}},
  \bibinfo {author} {\bibfnamefont {A.}~\bibnamefont {Ainla}}, \bibinfo
  {author} {\bibfnamefont {M.}~\bibnamefont {Zhao}}, \bibinfo {author}
  {\bibfnamefont {M.}~\bibnamefont {Baghbanzadeh}}, \bibinfo {author}
  {\bibfnamefont {V.~E.}\ \bibnamefont {Campbell}}, \bibinfo {author}
  {\bibfnamefont {K.}~\bibnamefont {Kang}}, \bibinfo {author} {\bibfnamefont
  {J.~M.}\ \bibnamefont {Fox}},\ and\ \bibinfo {author} {\bibfnamefont {G.~M.}\
  \bibnamefont {Whitesides}},\ }\bibfield  {title} {\bibinfo {title}
  {Autocatalytic, bistable, oscillatory networks of biologically relevant
  organic reactions},\ }\href {https://doi.org/10.1038/nature19776} {\bibfield
  {journal} {\bibinfo  {journal} {Nature}\ }\textbf {\bibinfo {volume} {537}},\
  \bibinfo {pages} {656} (\bibinfo {year} {2016})}\BibitemShut {NoStop}%
\bibitem [{\citenamefont {Lincoln}\ and\ \citenamefont
  {Joyce}(2009)}]{LincolnJoyce2009}%
  \BibitemOpen
  \bibfield  {author} {\bibinfo {author} {\bibfnamefont {T.~A.}\ \bibnamefont
  {Lincoln}}\ and\ \bibinfo {author} {\bibfnamefont {G.~F.}\ \bibnamefont
  {Joyce}},\ }\bibfield  {title} {\bibinfo {title} {Self-sustained replication
  of an {RNA} enzyme},\ }\href {https://doi.org/10.1126/science.1167856}
  {\bibfield  {journal} {\bibinfo  {journal} {Science}\ }\textbf {\bibinfo
  {volume} {323}},\ \bibinfo {pages} {1229} (\bibinfo {year}
  {2009})}\BibitemShut {NoStop}%
\bibitem [{\citenamefont {Vaidya}\ \emph {et~al.}(2012)\citenamefont {Vaidya},
  \citenamefont {Manapat}, \citenamefont {Chen}, \citenamefont {Xulvi-Brunet},
  \citenamefont {Hayden},\ and\ \citenamefont {Lehman}}]{Vaidya2012}%
  \BibitemOpen
  \bibfield  {author} {\bibinfo {author} {\bibfnamefont {N.}~\bibnamefont
  {Vaidya}}, \bibinfo {author} {\bibfnamefont {M.~L.}\ \bibnamefont {Manapat}},
  \bibinfo {author} {\bibfnamefont {I.~A.}\ \bibnamefont {Chen}}, \bibinfo
  {author} {\bibfnamefont {R.}~\bibnamefont {Xulvi-Brunet}}, \bibinfo {author}
  {\bibfnamefont {E.~J.}\ \bibnamefont {Hayden}},\ and\ \bibinfo {author}
  {\bibfnamefont {N.}~\bibnamefont {Lehman}},\ }\bibfield  {title} {\bibinfo
  {title} {Spontaneous network formation among cooperative {RNA} replicators},\
  }\href {https://doi.org/10.1038/nature11549} {\bibfield  {journal} {\bibinfo
  {journal} {Nature}\ }\textbf {\bibinfo {volume} {491}},\ \bibinfo {pages}
  {72} (\bibinfo {year} {2012})}\BibitemShut {NoStop}%
\bibitem [{\citenamefont {Adamala}\ and\ \citenamefont
  {Szostak}(2013)}]{AdamalaSzostak2013}%
  \BibitemOpen
  \bibfield  {author} {\bibinfo {author} {\bibfnamefont {K.}~\bibnamefont
  {Adamala}}\ and\ \bibinfo {author} {\bibfnamefont {J.~W.}\ \bibnamefont
  {Szostak}},\ }\bibfield  {title} {\bibinfo {title} {Competition between model
  protocells driven by an encapsulated catalyst},\ }\href
  {https://doi.org/10.1038/nchem.1650} {\bibfield  {journal} {\bibinfo
  {journal} {Nat. Chem.}\ }\textbf {\bibinfo {volume} {5}},\ \bibinfo {pages}
  {495} (\bibinfo {year} {2013})}\BibitemShut {NoStop}%
\bibitem [{\citenamefont {Menck}\ \emph {et~al.}(2013)\citenamefont {Menck},
  \citenamefont {Heitzig}, \citenamefont {Marwan},\ and\ \citenamefont
  {Kurths}}]{Menck2013}%
  \BibitemOpen
  \bibfield  {author} {\bibinfo {author} {\bibfnamefont {P.~J.}\ \bibnamefont
  {Menck}}, \bibinfo {author} {\bibfnamefont {J.}~\bibnamefont {Heitzig}},
  \bibinfo {author} {\bibfnamefont {N.}~\bibnamefont {Marwan}},\ and\ \bibinfo
  {author} {\bibfnamefont {J.}~\bibnamefont {Kurths}},\ }\bibfield  {title}
  {\bibinfo {title} {How basin stability complements the linear-stability
  paradigm},\ }\href {https://doi.org/10.1038/nphys2516} {\bibfield  {journal}
  {\bibinfo  {journal} {Nat. Phys.}\ }\textbf {\bibinfo {volume} {9}},\
  \bibinfo {pages} {89} (\bibinfo {year} {2013})}\BibitemShut {NoStop}%
\bibitem [{\citenamefont {E}\ and\ \citenamefont
  {Vanden-Eijnden}(2006)}]{EVandenEijnden2006}%
  \BibitemOpen
  \bibfield  {author} {\bibinfo {author} {\bibfnamefont {W.}~\bibnamefont {E}}\
  and\ \bibinfo {author} {\bibfnamefont {E.}~\bibnamefont {Vanden-Eijnden}},\
  }\bibfield  {title} {\bibinfo {title} {Towards a theory of transition
  paths},\ }\href {https://doi.org/10.1007/s10955-005-9003-9} {\bibfield
  {journal} {\bibinfo  {journal} {J. Stat. Phys.}\ }\textbf {\bibinfo {volume}
  {123}},\ \bibinfo {pages} {503} (\bibinfo {year} {2006})}\BibitemShut
  {NoStop}%
\bibitem [{\citenamefont {Metzner}\ \emph {et~al.}(2009)\citenamefont
  {Metzner}, \citenamefont {Sch{\"u}tte},\ and\ \citenamefont
  {Vanden-Eijnden}}]{Metzner2009}%
  \BibitemOpen
  \bibfield  {author} {\bibinfo {author} {\bibfnamefont {P.}~\bibnamefont
  {Metzner}}, \bibinfo {author} {\bibfnamefont {C.}~\bibnamefont
  {Sch{\"u}tte}},\ and\ \bibinfo {author} {\bibfnamefont {E.}~\bibnamefont
  {Vanden-Eijnden}},\ }\bibfield  {title} {\bibinfo {title} {Transition path
  theory for markov jump processes},\ }\href
  {https://doi.org/10.1137/070699500} {\bibfield  {journal} {\bibinfo
  {journal} {Multiscale Model. Simul.}\ }\textbf {\bibinfo {volume} {7}},\
  \bibinfo {pages} {1192} (\bibinfo {year} {2009})}\BibitemShut {NoStop}%
\bibitem [{\citenamefont {Be'er}\ \emph {et~al.}(2015)\citenamefont {Be'er},
  \citenamefont {Assaf},\ and\ \citenamefont {Meerson}}]{Beer2015}%
  \BibitemOpen
  \bibfield  {author} {\bibinfo {author} {\bibfnamefont {S.}~\bibnamefont
  {Be'er}}, \bibinfo {author} {\bibfnamefont {M.}~\bibnamefont {Assaf}},\ and\
  \bibinfo {author} {\bibfnamefont {B.}~\bibnamefont {Meerson}},\ }\bibfield
  {title} {\bibinfo {title} {Colonization of a territory by a stochastic
  population under a strong allee effect and a low immigration pressure},\
  }\href {https://doi.org/10.1103/PhysRevE.91.062126} {\bibfield  {journal}
  {\bibinfo  {journal} {Phys. Rev. E}\ }\textbf {\bibinfo {volume} {91}},\
  \bibinfo {pages} {062126} (\bibinfo {year} {2015})}\BibitemShut {NoStop}%
\bibitem [{\citenamefont {Kim}\ \emph {et~al.}(2014)\citenamefont {Kim},
  \citenamefont {Josi{\'c}},\ and\ \citenamefont
  {Bennett}}]{KimJosicBennett2014}%
  \BibitemOpen
  \bibfield  {author} {\bibinfo {author} {\bibfnamefont {J.~K.}\ \bibnamefont
  {Kim}}, \bibinfo {author} {\bibfnamefont {K.}~\bibnamefont {Josi{\'c}}},\
  and\ \bibinfo {author} {\bibfnamefont {M.~R.}\ \bibnamefont {Bennett}},\
  }\bibfield  {title} {\bibinfo {title} {The validity of quasi-steady-state
  approximations in discrete stochastic simulations},\ }\href
  {https://doi.org/10.1016/j.bpj.2014.06.012} {\bibfield  {journal} {\bibinfo
  {journal} {Biophys. J.}\ }\textbf {\bibinfo {volume} {107}},\ \bibinfo
  {pages} {783} (\bibinfo {year} {2014})}\BibitemShut {NoStop}%
\bibitem [{\citenamefont {Thieme}(2009)}]{Thieme2009}%
  \BibitemOpen
  \bibfield  {author} {\bibinfo {author} {\bibfnamefont {H.~R.}\ \bibnamefont
  {Thieme}},\ }\bibfield  {title} {\bibinfo {title} {Spectral bound and
  reproduction number for infinite-dimensional population structure and time
  heterogeneity},\ }\href {https://doi.org/10.1137/080732870} {\bibfield
  {journal} {\bibinfo  {journal} {SIAM J. Appl. Math.}\ }\textbf {\bibinfo
  {volume} {70}},\ \bibinfo {pages} {188} (\bibinfo {year} {2009})}\BibitemShut
  {NoStop}%
\bibitem [{\citenamefont {Farina}\ and\ \citenamefont
  {Rinaldi}(2000)}]{FarinaRinaldi2000}%
  \BibitemOpen
  \bibfield  {author} {\bibinfo {author} {\bibfnamefont {L.}~\bibnamefont
  {Farina}}\ and\ \bibinfo {author} {\bibfnamefont {S.}~\bibnamefont
  {Rinaldi}},\ }\href {https://doi.org/10.1002/9781118033029} {\emph {\bibinfo
  {title} {Positive Linear Systems: Theory and Applications}}}\ (\bibinfo
  {publisher} {Wiley},\ \bibinfo {address} {New York},\ \bibinfo {year}
  {2000})\BibitemShut {NoStop}%
\bibitem [{\citenamefont {Angeli}\ and\ \citenamefont
  {Sontag}(2004)}]{AngeliSontag2004}%
  \BibitemOpen
  \bibfield  {author} {\bibinfo {author} {\bibfnamefont {D.}~\bibnamefont
  {Angeli}}\ and\ \bibinfo {author} {\bibfnamefont {E.~D.}\ \bibnamefont
  {Sontag}},\ }\bibfield  {title} {\bibinfo {title} {Multi-stability in
  monotone input/output systems},\ }\href
  {https://doi.org/10.1016/j.sysconle.2003.08.003} {\bibfield  {journal}
  {\bibinfo  {journal} {Syst. Control Lett.}\ }\textbf {\bibinfo {volume}
  {51}},\ \bibinfo {pages} {185} (\bibinfo {year} {2004})}\BibitemShut
  {NoStop}%
\bibitem [{\citenamefont {Angeli}\ \emph {et~al.}(2004)\citenamefont {Angeli},
  \citenamefont {Ferrell},\ and\ \citenamefont {Sontag}}]{AngeliFerrell2004}%
  \BibitemOpen
  \bibfield  {author} {\bibinfo {author} {\bibfnamefont {D.}~\bibnamefont
  {Angeli}}, \bibinfo {author} {\bibfnamefont {J.~E.}\ \bibnamefont {Ferrell},
  \bibfnamefont {Jr.}},\ and\ \bibinfo {author} {\bibfnamefont {E.~D.}\
  \bibnamefont {Sontag}},\ }\bibfield  {title} {\bibinfo {title} {Detection of
  multistability, bifurcations, and hysteresis in a large class of biological
  positive-feedback systems},\ }\href {https://doi.org/10.1073/pnas.0308265100}
  {\bibfield  {journal} {\bibinfo  {journal} {Proc. Natl. Acad. Sci. U.S.A.}\
  }\textbf {\bibinfo {volume} {101}},\ \bibinfo {pages} {1822} (\bibinfo {year}
  {2004})}\BibitemShut {NoStop}%
\bibitem [{\citenamefont {Trefethen}\ \emph {et~al.}(1993)\citenamefont
  {Trefethen}, \citenamefont {Trefethen}, \citenamefont {Reddy},\ and\
  \citenamefont {Driscoll}}]{Trefethen1993}%
  \BibitemOpen
  \bibfield  {author} {\bibinfo {author} {\bibfnamefont {L.~N.}\ \bibnamefont
  {Trefethen}}, \bibinfo {author} {\bibfnamefont {A.~E.}\ \bibnamefont
  {Trefethen}}, \bibinfo {author} {\bibfnamefont {S.~C.}\ \bibnamefont
  {Reddy}},\ and\ \bibinfo {author} {\bibfnamefont {T.~A.}\ \bibnamefont
  {Driscoll}},\ }\bibfield  {title} {\bibinfo {title} {Hydrodynamic stability
  without eigenvalues},\ }\href {https://doi.org/10.1126/science.261.5121.578}
  {\bibfield  {journal} {\bibinfo  {journal} {Science}\ }\textbf {\bibinfo
  {volume} {261}},\ \bibinfo {pages} {578} (\bibinfo {year}
  {1993})}\BibitemShut {NoStop}%
\bibitem [{\citenamefont {Kurtz}(1972)}]{Kurtz1972}%
  \BibitemOpen
  \bibfield  {author} {\bibinfo {author} {\bibfnamefont {T.~G.}\ \bibnamefont
  {Kurtz}},\ }\bibfield  {title} {\bibinfo {title} {The relationship between
  stochastic and deterministic models for chemical reactions},\ }\href
  {https://doi.org/10.1063/1.1678692} {\bibfield  {journal} {\bibinfo
  {journal} {J. Chem. Phys.}\ }\textbf {\bibinfo {volume} {57}},\ \bibinfo
  {pages} {2976} (\bibinfo {year} {1972})}\BibitemShut {NoStop}%
\bibitem [{\citenamefont {Gillespie}(1977)}]{Gillespie1977}%
  \BibitemOpen
  \bibfield  {author} {\bibinfo {author} {\bibfnamefont {D.~T.}\ \bibnamefont
  {Gillespie}},\ }\bibfield  {title} {\bibinfo {title} {Exact stochastic
  simulation of coupled chemical reactions},\ }\href
  {https://doi.org/10.1021/j100540a008} {\bibfield  {journal} {\bibinfo
  {journal} {J. Phys. Chem.}\ }\textbf {\bibinfo {volume} {81}},\ \bibinfo
  {pages} {2340} (\bibinfo {year} {1977})}\BibitemShut {NoStop}%
\bibitem [{\citenamefont {Gillespie}(2000)}]{Gillespie2000}%
  \BibitemOpen
  \bibfield  {author} {\bibinfo {author} {\bibfnamefont {D.~T.}\ \bibnamefont
  {Gillespie}},\ }\bibfield  {title} {\bibinfo {title} {The chemical langevin
  equation},\ }\href {https://doi.org/10.1063/1.481811} {\bibfield  {journal}
  {\bibinfo  {journal} {J. Chem. Phys.}\ }\textbf {\bibinfo {volume} {113}},\
  \bibinfo {pages} {297} (\bibinfo {year} {2000})}\BibitemShut {NoStop}%
\bibitem [{\citenamefont {Freidlin}\ and\ \citenamefont
  {Wentzell}(2012)}]{FreidlinWentzell2012}%
  \BibitemOpen
  \bibfield  {author} {\bibinfo {author} {\bibfnamefont {M.~I.}\ \bibnamefont
  {Freidlin}}\ and\ \bibinfo {author} {\bibfnamefont {A.~D.}\ \bibnamefont
  {Wentzell}},\ }\href {https://doi.org/10.1007/978-3-642-25847-3} {\emph
  {\bibinfo {title} {Random Perturbations of Dynamical Systems}}},\ \bibinfo
  {edition} {3rd}\ ed.\ (\bibinfo  {publisher} {Springer},\ \bibinfo {address}
  {Heidelberg},\ \bibinfo {year} {2012})\BibitemShut {NoStop}%
\bibitem [{\citenamefont {Dykman}\ \emph {et~al.}(1994)\citenamefont {Dykman},
  \citenamefont {Mori}, \citenamefont {Ross},\ and\ \citenamefont
  {Hunt}}]{Dykman1994}%
  \BibitemOpen
  \bibfield  {author} {\bibinfo {author} {\bibfnamefont {M.~I.}\ \bibnamefont
  {Dykman}}, \bibinfo {author} {\bibfnamefont {E.}~\bibnamefont {Mori}},
  \bibinfo {author} {\bibfnamefont {J.}~\bibnamefont {Ross}},\ and\ \bibinfo
  {author} {\bibfnamefont {P.~M.}\ \bibnamefont {Hunt}},\ }\bibfield  {title}
  {\bibinfo {title} {Large fluctuations and optimal paths in chemical
  kinetics},\ }\href {https://doi.org/10.1063/1.467139} {\bibfield  {journal}
  {\bibinfo  {journal} {J. Chem. Phys.}\ }\textbf {\bibinfo {volume} {100}},\
  \bibinfo {pages} {5735} (\bibinfo {year} {1994})}\BibitemShut {NoStop}%
\bibitem [{\citenamefont {Horn}\ and\ \citenamefont
  {Jackson}(1972)}]{HornJackson1972}%
  \BibitemOpen
  \bibfield  {author} {\bibinfo {author} {\bibfnamefont {F.}~\bibnamefont
  {Horn}}\ and\ \bibinfo {author} {\bibfnamefont {R.}~\bibnamefont {Jackson}},\
  }\bibfield  {title} {\bibinfo {title} {General mass action kinetics},\ }\href
  {https://doi.org/10.1007/BF00251225} {\bibfield  {journal} {\bibinfo
  {journal} {Arch. Ration. Mech. Anal.}\ }\textbf {\bibinfo {volume} {47}},\
  \bibinfo {pages} {81} (\bibinfo {year} {1972})}\BibitemShut {NoStop}%
\bibitem [{\citenamefont {Rao}\ and\ \citenamefont
  {Esposito}(2016)}]{RaoEsposito2016}%
  \BibitemOpen
  \bibfield  {author} {\bibinfo {author} {\bibfnamefont {R.}~\bibnamefont
  {Rao}}\ and\ \bibinfo {author} {\bibfnamefont {M.}~\bibnamefont {Esposito}},\
  }\bibfield  {title} {\bibinfo {title} {Nonequilibrium thermodynamics of
  chemical reaction networks: Wisdom from stochastic thermodynamics},\ }\href
  {https://doi.org/10.1103/PhysRevX.6.041064} {\bibfield  {journal} {\bibinfo
  {journal} {Phys. Rev. X}\ }\textbf {\bibinfo {volume} {6}},\ \bibinfo {pages}
  {041064} (\bibinfo {year} {2016})}\BibitemShut {NoStop}%
\bibitem [{\citenamefont {Catling}\ \emph {et~al.}(2018)\citenamefont
  {Catling}, \citenamefont {Krissansen-Totton}, \citenamefont {Kiang},
  \citenamefont {Crisp}, \citenamefont {Robinson}, \citenamefont {DasSarma},
  \citenamefont {Rushby}, \citenamefont {Del~Genio}, \citenamefont {Bains},\
  and\ \citenamefont {Domagal-Goldman}}]{Catling2018}%
  \BibitemOpen
  \bibfield  {author} {\bibinfo {author} {\bibfnamefont {D.~C.}\ \bibnamefont
  {Catling}}, \bibinfo {author} {\bibfnamefont {J.}~\bibnamefont
  {Krissansen-Totton}}, \bibinfo {author} {\bibfnamefont {N.~Y.}\ \bibnamefont
  {Kiang}}, \bibinfo {author} {\bibfnamefont {D.}~\bibnamefont {Crisp}},
  \bibinfo {author} {\bibfnamefont {T.~D.}\ \bibnamefont {Robinson}}, \bibinfo
  {author} {\bibfnamefont {S.}~\bibnamefont {DasSarma}}, \bibinfo {author}
  {\bibfnamefont {A.~J.}\ \bibnamefont {Rushby}}, \bibinfo {author}
  {\bibfnamefont {A.}~\bibnamefont {Del~Genio}}, \bibinfo {author}
  {\bibfnamefont {W.}~\bibnamefont {Bains}},\ and\ \bibinfo {author}
  {\bibfnamefont {S.}~\bibnamefont {Domagal-Goldman}},\ }\bibfield  {title}
  {\bibinfo {title} {Exoplanet biosignatures: A framework for their
  assessment},\ }\href {https://doi.org/10.1089/ast.2017.1737} {\bibfield
  {journal} {\bibinfo  {journal} {Astrobiology}\ }\textbf {\bibinfo {volume}
  {18}},\ \bibinfo {pages} {709} (\bibinfo {year} {2018})}\BibitemShut
  {NoStop}%
\bibitem [{\citenamefont {Kuznetsov}(2004)}]{Kuznetsov2004}%
  \BibitemOpen
  \bibfield  {author} {\bibinfo {author} {\bibfnamefont {Y.~A.}\ \bibnamefont
  {Kuznetsov}},\ }\href {https://doi.org/10.1007/978-1-4757-3978-7} {\emph
  {\bibinfo {title} {Elements of Applied Bifurcation Theory}}},\ \bibinfo
  {edition} {3rd}\ ed.\ (\bibinfo  {publisher} {Springer},\ \bibinfo {address}
  {New York},\ \bibinfo {year} {2004})\BibitemShut {NoStop}%
\bibitem [{\citenamefont {Strogatz}(2015)}]{Strogatz2015}%
  \BibitemOpen
  \bibfield  {author} {\bibinfo {author} {\bibfnamefont {S.~H.}\ \bibnamefont
  {Strogatz}},\ }\href@noop {} {\emph {\bibinfo {title} {Nonlinear Dynamics and
  Chaos}}},\ \bibinfo {edition} {2nd}\ ed.\ (\bibinfo  {publisher} {Westview
  Press},\ \bibinfo {address} {Boulder},\ \bibinfo {year} {2015})\BibitemShut
  {NoStop}%
\bibitem [{\citenamefont {Segel}\ and\ \citenamefont
  {Slemrod}(1989)}]{SegelSlemrod1989}%
  \BibitemOpen
  \bibfield  {author} {\bibinfo {author} {\bibfnamefont {L.~A.}\ \bibnamefont
  {Segel}}\ and\ \bibinfo {author} {\bibfnamefont {M.}~\bibnamefont
  {Slemrod}},\ }\bibfield  {title} {\bibinfo {title} {The quasi-steady-state
  assumption: A case study in perturbation},\ }\href
  {https://doi.org/10.1137/1031091} {\bibfield  {journal} {\bibinfo  {journal}
  {SIAM Rev.}\ }\textbf {\bibinfo {volume} {31}},\ \bibinfo {pages} {446}
  (\bibinfo {year} {1989})}\BibitemShut {NoStop}%
\end{thebibliography}

%

\end{document}